# Mineralogical Characterization of Baptistina Asteroid Family: Implications for K/T Impactor Source

Vishnu Reddy[1,2]
Department of Space Studies, Room 520, Box 9008, University of North Dakota, Grand Forks, ND 58202, USA.
Email: reddy@space.edu

Jorge M. Carvano
Observatório Nacional (COAA), rua Gal. José Cristino 77, São Cristóvão, CEP20921-400 Rio de Janeiro RJ, Brazil.

Daniela Lazzaro
Observatório Nacional (COAA), rua Gal. José Cristino 77, São Cristóvão, CEP20921-400 Rio de Janeiro RJ, Brazil.

Tatiana A. Michtchenko
Institute for Astronomy, Geophysics, and Atmospheric Sciences, University of São Paulo, Brazil.

Michael J. Gaffey[1]
Department of Space Studies, Room 518, Box 9008, University of North Dakota, Grand Forks, ND58202, USA.

Michael S. Kelley[1,3]
Department of Geology and Geography, Box 8149, Georgia Southern University, Statesboro, GA 30460, USA.

Thais Mothé Diniz[1]
Observatório do Valongo, Federal University of Rio de Janeiro, Rio de Janeiro, RJ, Brazil.

Alvaro Alvarez Candal
, Alonso de Córdova 3107, Vitacura, Casilla 19001, Santiago 19, Chile.

Nicholas A. Moskovitz[1]
Carnegie Institution of Washington, Department of Terrestrial Magnetism, 5241, Broad Branch Road, Washington, DC 20008, USA

Edward A. Cloutis
Department of Geography, University of Winnipeg, 515 Portage Avenue, Winnipeg, Manitoba, Canada R3B 2E9

Erin L. Ryan
Department of Astronomy, University of Minnesota, MN 55455, USA

Pages: 62
Tables: 6
Figures: 15

**Proposed Running Head:** Baptistina Asteroid Family

**Editorial correspondence to:**
Vishnu Reddy
Department of Space Studies
Room 520, Box 9008
University of North Dakota
Grand Forks, ND 58202
(808) 342-8932 (voice)
(701) 777-3711 (fax)
reddy@space.edu

**Abstract**

Bottke et al. (2007) linked the catastrophic formation of Baptistina Asteroid Family (BAF) to the K/T impact event. This linkage was based on dynamical and compositional evidence, which suggested the impactor had a composition similar to CM2 carbonaceous chondrites. However, our recent study (Reddy et al. 2009) suggests that the composition of (298) Baptistina is similar to LL-type ordinary chondrites rather than CM2 carbonaceous chondrites. This rules out any possibility of it being related to the source of the K/T impactor, if the impactor was of CM-type composition. Mineralogical study of asteroids in the vicinity of BAF has revealed a plethora of compositional types suggesting a complex formation and evolution environment. A detailed compositional analysis of 16 asteroids suggests several distinct surface assemblages including ordinary chondrites (Gaffey SIV subtype), primitive achondrites (Gaffey SIII subtype), basaltic achondrites (Gaffey SVII subtype and V-type), and a carbonaceous chondrite. Based on our mineralogical analysis we conclude that (298) Baptistina is similar to ordinary chondrites (LL-type) based on olivine and pyroxene mineralogy and moderate albedo. S-type and V-type in and around the vicinity of BAF we characterized show mineralogical affinity to (8) Flora and (4) Vesta and could be part of their families. Smaller BAF asteroids with lower SNR spectra showing only a 'single' band are compositionally similar to (298) Baptistina and L/LL chondrites. It is unclear at this point why the silicate absorption bands in spectra of asteroids with formal family definition seem suppressed relative to background population, despite having similar mineralogy.

# 1. Introduction

Asteroid families are assumed to be the remnants of catastrophic collisions among asteroids and are identified as clusters in orbit element space. In this sense, asteroid families are important laboratories for studying the collisional evolution in the Main Belt as well as the structure and interior of their parent bodies. Since the pioneering work by Hirayama (1918), relatively few families have been mutually recognized by the variety of existing analytical methods. The use of different clustering techniques and proper element computation methods has led to discrepancies in asteroid family identification. A bigger problem is that the inventory of known asteroids remains incomplete. With every new analysis the number of objects nearly doubles compared to previous studies leading to strengthening or weakening the identification of some clusters, and leading to the identification of new families while others are discarded (Zappalla et al. 1990, 1994, 1995; Mothe-Diniz et al. 2005).

A genetic family, as initially defined by Farinella et al. (1992), is a cluster of asteroids whose physical properties are compatible with the fragmentation of a parent body. Determining the mineralogy of the family members or, in the absence of mineralogical information, the taxonomic classification for a significant number of family members can test the genetic viability of putative asteroid family. The members of a genetic family need to share a plausible cosmochemical origin (Chapman et al. 1989). Taxonomic classification has limited correlation to surface mineralogy of an asteroid and therefore provides a weaker test of genetic relationships. In the last decade several studies have been performed on family identification from a dynamical/clustering point of view and on the physical characterization of its members, either from taxonomy or mineralogy.

These issues have been extensively reviewed in the Asteroid III book (e.g., Bendjoya & Zappalà 2002; Zappalà et al. 2002, and Cellino et al. 2002).

The Baptistina Asteroid Family (BAF), at the inner border of the Main Belt (~2.25 AU), was identified only recently in the work by Mothé-Diniz et al. (2005), using the Hierachical Clustering Method on a sample of more than 120,000 asteroids with computed proper elements. The high concentration of asteroids in this region was previously associated with the Flora families, or with several distinct smaller families (Hirayama 1919; Williams 1992; Zappalà et al. 1990, 1994, 1995; Nesvorný et al. 2005). Mothé-Diniz et al. (2005), however, identified just one family in the region, with (298) Baptistina as its largest member, and some clumps around Flora.

In their work, these authors also used the available taxonomic information on the family members in order to study their mineralogical diversity and their relationship with the background objects. An interesting aspect of the BAF is the taxonomic diversity observed by Mothé-Diniz et al. (2005). Of the few members (8) with visible spectra, they found two S-, one Xc-, one X-, one C-, one A-, one V-, and one L-type in the family. According to the authors, this distribution of taxonomies was very similar to that of the background, making the family indistinguishable, from a taxonomic point of view. On the other hand, they also suggested that this distribution was compatible with that expected from the breakup of a differentiated parent body.

The Baptistina family was again identified by Bottke et al. (2007) as a cluster in proper elements and showed a peculiar concentration of C- or X-type asteroids in a region dominated by S-types. This was based on the analysis of broadband colors from the Sloan Digital Survey Moving Object Catalog (SDSS MOC) (Izevic et al. 2001), and was later confirmed with an updated version of the SDSS MOC (Parker et al. 2008). These authors introduced the Sloan colors in their statistical search for families and identified a Baptistina family, 'buried' within the Flora family.

In 2007, Bottke et al. considering the spread in semi-major axis of BAF members and making assumptions about the albedo, computed the age of the family to be $160^{+30}_{-20}$ Myr. According to the authors this age would set it as the "most likely source of the Chicxulub impactor that produced the Cretaceous/Tertiary (K/T) mass extinction event 65 Mys ago." Moreover, the C- and X-type colors of (298) Baptistina and most of the family members (from Sloan colors), would be compatible with a parent body ~170 km in diameter and with a CM2 carbonaceous chondrite-like composition, similar to that of a fossil meteorite found in K/T boundary sediments (Kyte 1998, Shukolyukov & Lugmair 1998, Trinquer et al. 2006).

In an effort to test Bottke et al. (2007) hypothesis we wanted to answer a simple question: Does (298) Baptistina have a composition/albedo similar to CM2 carbonaceous chondrites? (298) Baptistina is the largest member of the Baptistina Asteroid family and was classified as an X (Tholen) or Xc (Bus) taxonomic type by the S3OS2 survey (Lazzaro et al. 2004). The X taxonomic class under the Tholen system includes E, M, and P types and is degenerate unless albedo information is available. This was reinforced by Carvano et al. (2003), who showed that an X-type spectrum is compatible with that of meteorites ranging from CM2 carbonaceous chondrites to pallasites.

In 2008, we observed (298) Baptistina to mineralogically characterize the object and showed the presence of a well-resolved absorption band at $1.0 \pm 0.01$ μm and a weaker band at $2.0 \pm 0.2$ μm (Reddy et al. 2009). Based on these absorption bands we determined that the surface composition of Baptistina is similar to that of S-type asteroids with mineralogy similar to LL chondrites. This would suggest that the surface composition of (298) Baptistina is incompatible with C/X taxonomic type and CM2 meteorite analog. Further confirming this result, Carvano & Lazzaro (2010) derived a visible geometric albedo for (298) Baptistina of 0.347. This albedo, derived from mid-infrared observations, again rules out any possibility of the asteroid being linked to CM2 carbonaceous chondrites, which have an average albedo of ~0.04 (Gaffey, 1976). Based on the mineralogy and albedo (298) Baptistina is unlikely to be related to the K/T impactor source assuming its composition was similar to CM2 carbonaceous chondrite.

The Baptistina Asteroid Family was so designated because of its lowest numbered and probably the largest member, (298) Baptistina. It is possible that Baptistina is simply an interloper in this region, and not representative of the family. But there was no detailed mineralogical information about the interlopers in the vicinity of BAF. So answering this question would require additional observations of smaller BAF members. To understand the true nature of the BAF and verify the link between its member asteroids we performed a detailed mineralogical analysis using NIR spectroscopy. Our goal was to answer some simple questions, 1) What is the real composition of (298) Baptistina? 2) Is (298) an interloper in its own family if it is not a dark C-type? 3) Do Sloan colors of other BAF members (Xc taxonomy) translate to a CM2 carbonaceous chondrite composition?

## 2. Family definition and target selection

In order to select the targets for the spectroscopic analysis of the family it is necessary to first choose a definition for the family. The most used tool to define asteroid families is the Hierarchical Clustering Method (HCM), introduced by Zappalà et al. (1990). However, the list of the formal members of a dynamical family defined by this method depends on the adopted choice of the cutoff velocity. As noted earlier, Mothé-Diniz et al. (2005) first defined the Baptistina family in an effort to detect asteroid families in the whole Main Belt using an updated catalog of proper elements. As such, the authors divided the Main Belt in three regions - inner, intermediate and outer belt - and adopted fixed cutoff values for each region as the average minimum distance between all neighboring orbits in each region. Baptistina was thus defined using the value for the inner Belt (V=57m/s), with 543 formal members. Bottke et al. (2007) on the other hand adopted initially V=53m/s on the basis that this cutoff yields a family which seems compatible with what is expected of an Yarkovsky-evolved family. A problem here is that Baptistina is located in a very dynamically complex region, which complicates a meaningful and comprehensive formal membership definition. On the following, we discuss briefly the main dynamical aspects of the Baptistina region and explore the effect of different cutoff to the family definition.

*2.1 Dynamical Analysis*

The dynamics of the region around (298) Baptistina was studied in detail by Michtchenko et al. (2010). Here we briefly summarize the main features of this dynamical neighborhood delimited by a = 2.15AU-2.35AU, e = 0.12-0.18 and I<10$^o$, where a, e and I are the proper semi major axis, eccentricity and inclination, respectively.

Figure 1 shows that the Baptistina group is located in a dynamically active region. The region around Baptistina is intersected by numerous mean motion and secular resonances in proper elements (a, I)-space. The construction of dynamical maps is described in detail in Michtchenko et al. (2002). The mean-motion resonances (MMR) shown by vertical gray lines in Fig. 1 are not very strong in this region, because of high order of those involving Jupiter--Saturn system that is acting from the right-hand side, and the very small mass of Mars, which is acting from the left-hand side of the inner belt. Many objects are found to be involved in the mean-motion resonances, mainly in the 7J/2A MMR with Jupiter and the 1A/2M MMR with Mars. As seen in Fig. 1, several members of the Baptistina group evolve inside the 7J/2A MMR. In contrast, secular resonances (SR) are strong and densely cover the Baptistina region. The most important is the linear $\nu_6$–SR, which forms an upper boundary of the low-inclination population of the region. There are also a huge number of low order non-linear secular resonances, which are associated to the zones of instability in Fig. 1. The effects of non-linear SR are important for the transport of small objects. Non-linear secular resonances, interacting with the Yarkovsky dissipation, excite the asteroid eccentricities and inclination and provoke their slow diffusion of the objects in to high eccentricity and inclination regions (Carruba et al. 2005). It is interesting to observe that (298) Baptistina itself evolves deeply inside the $(2\nu_6 + \nu_{16})$ –SR, with period of ~6 Myr, and is very close to the $(\nu_{16} + \nu_5 - \nu_4)$ –SR, with period of ~14 Myr. Moreover, the size of Baptistina (~12.4 km) is sufficiently small to experience Yarkovsky-driven migration. These facts suggest that (298) Baptistina has probably been transported to its current position in proper elements space. The dissipative nature of this process makes impossible any conjecture about the original position of the object and the instant of the capture inside the $(2n_6 + \nu_{16})$ –SR.

The superposition of many, even weak, resonances produces efficient mechanisms to disperse the population in the region of overlapping resonances. As shown in (Michtchenko et al. 2010), there are several signs that the objects from the Baptistina region have experienced the effects of these mechanisms. All these considerations allow even to question, from a pure dynamical perspective, whether the Baptistina group is a real family or it is just an agglomeration of the objects produced by dynamical effects of present resonances.

*2.2 HCM analysis of the neighborhood of (298) Baptistina*

The velocity cutoff, used by the HCM method, is a function of the number of objects present in the local background, which also depends on the number of asteroids available in the constantly updated database of asteroid proper elements of Milani & Knežević (1994). Usually, the cutoff is set to the value for which the family coalesces with the local background, enclosing members of other known asteroid families. If there are no other families in the region, the choice of the cutoff is ambiguous. To illustrate this difficulty, we apply HCM to assess the Baptistina hypothetical family. Our results are shown in Fig. 2, where we plot the number of the possible family members as a function of velocity cutoff (top panel). In the bottom panel, we plot the normalized number of family members that are added to the family when the cutoff is increased by a given step. The cluster originated by (298) Baptistina appears at cutoff of 42 m/s, increases twofold at 46 m/s and clearly merges with the whole population of the region at 61 m/s. For intermediate cutoff values, we cannot identify any event, which could indicate that the family coalesces with the local background. Since there is no justification for the choice of a specific value of velocity cutoff in the range between 42 m/s and 61 m/s, we have chosen three different values, 45 m/s, 50 m/s and 55 m/s, and calculated the corresponding families applying HCM.

The clusters of the objects obtained for three values of the velocity cutoff are shown in Fig. 3. Two different planes are used to expose the putative family: the (a,I)-plane of the proper semi major axis and inclination (left column) and the (g,g+s)-plane of the proper perihelion, g, and node, s, frequencies (right column). The first subspace is widely used to show an asteroid family, while the second one is of a special interest for our approach. This is because the secular resonances, linear and nonlinear, are easily identified in the frequency space, in contrast with the proper elements space, and the relative to SR positions of the objects are very precise in this space.

The density (the number of asteroids inside a cell) of the objects in the proper frequencies subspace is shown by different colors: the green corresponds to minimal values (<10) and red to maximal values (>60). The evolution of the family with the increasing velocity cutoff value shows an unusual behavior. At small value of cutoff 45 m/s (Fig. 3, top row), the detected family is extended along the strong $(2\nu_6 + \nu_{16})$-SR and in the opposite direction with respect to the highest density: almost all members in this sample are influenced by the resonance in which the Baptistina is involved. This behavior of the low cutoff group of objects contrasts with a typical HCM outcome for conventional families, where a family grows around the parent body. For the larger cutoff value of 50 m/s, the group is extended covering the outer lobe (Fig. 3, middle row). And, for the cutoff value of 55 m/s (Fig. 3, bottom row), the possible family fills almost whole region, extending up to the (8) Flora (magenta symbol).

The above-described features show that the complicated dynamical structure of the region makes the identification of the family very difficult. For instance, the Baptistina group could be just an accumulation of objects in the intricate region of the overlapping of secular and mean-motion resonances. This assumption is consistent with the analysis of the distribution of the objects in the frequency space shown in Fig. 3 (right column). The neighborhood of the object shows elevated density, which curiously follows the important bands of the secular resonances, while (298) Baptistina itself is evolving along two bands. In addition, the contour of the asteroid agglomeration around Baptistina contrasts with the shapes of the other families, characterized by low dispersion in inclinations. The wide dispersion of the Baptistina group is a curious feature, suggesting different dynamical mechanisms acting in the region.

We apply then the HCM to the database of BAF members with known taxonomic classes as determined from SDSS colors (Carvano et al. 2010). Two samples were used: one with S-type objects and other with C-/X-type objects (both already reported in Section 4). The S-type sample consists of about 4,000 objects. Applying HCM to this sample, we detected no clusters for velocity cutoff values up to 160 m/s. On the other hand, we have detected a family in the C-/X-type sample, which consists of about 2,000 objects. The sample is shown on the (a, I)—plane in Fig. 4 by black dots, where the main MMRs (vertical lines) and the strong linear and non-linear SRs in the region are sketched. The family detected within the C-/X sample is shown by cyan symbols in Fig. 4. This family was calculated with a velocity cutoff of 81 m/s, which may be slightly elevated due to low number statistics. The family consists of 105 members and its location coincides with the outer lobe of the highest density, although the biggest member (289) Baptistina is far from the main bulk of the objects. Thus, a dynamical family is detected only for object with C/X taxonomy. This is equivalent to the results of Parker et al. (2008), whose family determination scheme uses both SDSS colors and proper elements to define a metric. Indeed, Baptistina is one of the few families in Parker et al. (2008), which could only be detected through SDSS colors.

The conclusion here is that the existence of Baptistina as a collisional family can be asserted only due to the marked difference in color from the densest concentration of objects around (298) Baptistina and its immediate vincinity. Since there's no obvious way to produce such color difference by dynamics alone, it follows that the Baptistina family must be the remnants of a true collisional family, even if it cannot be robustly defined by hierarchical clustering methods. Therefore a robust list of formal members for the Baptistina family should in principle also consider SDSS colors. Such family would be by construction spectroscopically homogeneous, at least on visible wavelengths.

Alternatively, one could use an approach similar to Bottke at al. (2007) and try to constrain family membership through Yarkovsky-driven dispersion. This however also implicitly assumes that the family members have homogeneous surface properties. Instead, in order to avoid a family definition that would preclude by construction the possibility of a diffentiated parent body, we chose to select our targets from a purely dynamical definition, using Vcutoff =53 m/s, taking care to select targets from both the more dense, family core, dominated by objects with C/X-type colors), and also from the family outskirts, dominated by objects with S-type colors. This would allow constraining the mineralogical diversity between the family core and its immediate vicinity. The final aim here is to use the derived mineralogy to assess the genetic relationship of the asteroids in the sample and then use them to interpret the distribution of color in the family.

## 3. Observations and Data Reduction

The visible spectra (0.49-0.92 μm) of 20 asteroids in the dynamical region near (298) Baptistina were acquired using the 1.5-m telescope at the European Southern Observatory, La Silla, Chile, and the 3.58 m Telescopio Nazionale Galileo at the Roque de Los Muchachos Observatory, La Palma, Canary Islands, Spain. The observations at ESO were carried out during four observing runs in November 1996 and January 1997, as part of the survey S3OS2 (Lazzaro et al. 2004), and in March and September 2002. We used the 1.5-m telescope equipped with a Boller and Chivens spectrograph and a 2048 x 2048 pixel CCD with a readout noise of 7[e–$rms$] and a square pixel of 15 μm. A grating of 225 gr/mm with a dispersion of 330 Å/mm in the first order was used. The configuration resulted in a useful spectral range of 4900 Å < λ < 9200 Å with a FWHM of 10 Å. The observational circumstances for the 14 asteroids observed at La Silla are given in Table 1. These data were reduced using methods described in Lazzaro et al. (2004).

Near-IR (0.7-2.5 μm) spectral observations of BAF members were conducted using the low-resolution SpeX instrument in prism mode (Rayner et al. 2003) on the NASA IRTF, Mauna Kea, Hawai'i between February 2008 and November 2009. Apart from BAF members, local standard stars and solar analog star observations were also performed to correct for telluric and solar continuum, respectively. Detailed description of the observing protocol is presented in Reddy (2009). Observational circumstances for SpeX data are shown in Table 2.

SpeX prism data were processed using two different data reduction protocols to cross-calibrate any variations in the spectral band parameters. The first method was developed by Gaffey and involved using a combination of Unix-based IRAF and Windows-based SpecPR software routines. Detailed description of this method can be found in Reddy (2009), Abell (2003), and Hardersen (2003). The data also were reduced using the IDL-based Spextool provided by the NASA IRTF (Cushing et al. 2004). Analysis of the data to determine band centers and areas was done using SpecPR, based on the protocols discussed by Cloutis et al. (1986), Gaffey et al. (2002), and Gaffey (2003, 2005).

**4. Is (298) Baptistina an interloper in its own family?**

Reddy et al. (2009) suggested that the composition of (298) Baptistina was similar to oridinary chondrites, most probably LL-type. But they stayed short of mineralogically characterizing the object. In order to verify if Baptistina is an interloper in its own family (assumed to be dominated by dark C-type from SDSS colors), first the composition of Baptistina must be constrained, then the next step would be to compare this composition with that of the assumed family composition (CM2 carbonaceous chondrites). Since Baptistina NIR spectrum shows Band I and II absorption bands we have applied Dunn et al. (2010) equations to derived its olivine and pyroxene ratios using:

Olv/(Olv+Pyx) = -0.242 * BAR + 0.728 (Eq. 1)

The RMS error for the spectrally derived values of olivine is 0.03 (Dunn et al. 2010). The chemistry of the olivine (fayalite or Fa) and mean pyroxene chemistry (ferrosilite or Fs) were also estimated using equations derived by Dunn et al. (2010)

Fa = -1284.9 * (Band I Center)$^2$ + 2656.5 * (Band I Center) – 1342.2 (Eq. 2)

Fs = -879.1 * (Band I Center)$^2$ + 1824.9 * (Band I Center) – 921.7 (Eq. 3)

(298) Baptistina (Fig. 5), which plots in the S(IV) region of the S-asteroid

subtypes plot (Fig. 6) close to the olivine-pyroxene mixing line. Gaffey et al. (1993) and Dunn et al. (2010) note that LL-chondrites fall in this region of the S(IV) region on Figure 6. Baptistina also falls within the LL-chondrite region on a Band I vs. Band II plot. The olivine (0.70) and pyroxene (0.30) percentages for Baptistina are within the olivine/pyroxene ratios for LL-chondrite olivine (0.60-0.70) and pyroxene (0.40-0.30) (Dunn et al. 2010). This value is consistent with the olivine/pyroxene ratio of 0.80:0.20±0.10 derived from the Cloutis et al. (1986) calibration and first reported by Reddy et al. (2009). The Fa and Fs content for Baptistina ($Fa_{29}$ and $Fs_{24}$) are also consistent with LL chondrites, which have a Fa range between $Fa_{27-33}$ and a Fs range between $Fs_{23-27}$ (Dunn et al. 2010).

However, as first reported by Reddy et al. (2009), both spectral bands of Baptistina are shallower (Band I depth ~7%) when compared to a typical S(IV) asteroid (12%). They suggested a range of possible explanations for this including the presence of opaques such as carbon or metal. In light of new albedo measurements ($p_v$ = 0.347) by (Carvano and Lazzaro, 2010), one can rule out carbon as the suppressing agent as that would also decrease the albedo. LL chondrites typically have albedos between 0.23-0.37 (Gaffey, 1976), which is consistent with Carvano & Lazzaro (2010) measurement for Baptistina. Based on its composition and albedo (298) Baptistina would be an interloper in its own family, assuming the family is of CM2 composition as suggested by Bottke et al. (2007). However, the question remains if there are other asteroids in or around BAF that have a composition similar to (298) Baptistina. In an effort to answer this question this, we mineralogically characterized several asteroids in the Baptistina region, which displayed Band I and II absorption bands (S- and V-types).

*5.1 S-type asteroids*

Figure 6 shows the sub-classification of S- and V-type BAF asteroids based on Gaffey et al. (1993) classification using their Band I center and Band Area Ratio (BAR). Two distinct groups emerge from this plot, one is the S(III)-S(IV) group (Fig. 5 and 7) and the other is the S(VII)-Basaltic Achondrite group (Fig. 8). Three asteroids, (3260) Vizbor, (6266) Letzel, and (13154) Petermrva, fall under the S(III) subtype on Fig. 7. These three objects have a larger percentage of high-Ca pyroxene than the S(IV) asteroids, which causes their band parameters to be offset from the olivine-orthopyroxene mixing line. We calculated olivine-pyroxene ratios for the S(III) subtypes using Eq. 1 which suggests a decrease in olivine content and an increase in pyroxene compared to the S(IV) subtypes (Table 3). Along with (298) Baptistina three other asteroids from the BAF can be classified as S(IV) subtypes, (1126) Otero, (1365) Henyey, and (2545) Verbiest. Mineralogical analysis of the S(IV) subtypes using Eq. 2 and 3 suggests that they belong to either L- or LL-chondrites based on their pyroxene chemistry.

Exploring the compositional link between the Flora clan and the S(III), S(IV) BAF can help to address whether the observed S-types are part of the background S-asteroid population. To accomplish this we compared the composition of the S(III) and S(IV) asteroids to (8) Flora, which is the largest member of the S-type Flora clan, that dominates the background population in this region. Flora plots in the border between the S(IV) and the upper S(III) zone close (3260) Vizbor (Fig. 6). Gaffey (1984) interpreted (8) Flora as a differentiated surface assemblage based on relative variations in the mineralogical / spectral parameters which had the wrong slope to be chondritic (undifferentiated). Flora's mineralogy was estimated to be, olivine/pyroxene ratio (0.74:0.26), metal content (red slope) and pyroxene chemistry ($Fs_{70\pm5}$).

Using Flora's band parameters we have reanalyzed Flora's mineralogy using new calibrations for ordinary chondrites (Dunn et al. 2010). While the Dunn et al. (2010) calibration can only be applied to S(IV) subtypes, Flora sits on the edge of the S(III-IV) region (Fig. 6) and thus the use of the new calibration is justified. Flora's derived olivine/pyroxene ratio is (0.65:0.35), olivine chemistry $Fa_{29}$ and pyroxene chemistry $Fs_{24}$. These values are similar to Baptistina and Petermrva except for the olivine/pyroxene ratio, which is intermediate for Flora (probably due to higher calcic pyroxene). The olivine and pyroxene chemistries of Flora are inconsistent with a differentiated asteroid (Gaffey, 1984), although low-degree partial melting cannot be ruled out. While the olivine/pyroxene ratio and chemistries of (8) Flora and the S(III) and S(IV) asteroids are broadly consistent, the suppression of absorption bands, especially for Baptistina, might be due to the presence of metal on the surface. The mineralogy, mineral abundance and chemistry of the S(III) and S(IV) subtypes in the vicinity of Baptistina strongly support the argument that a majority of them are consistent with the background S-asteroid population, namely the Flora clan.

*5.2 V-type asteroids*

Two asteroids in the sample (13480) Potapov and (24245) Ezratty (Fig. 8) plot in the S(VII) zone of Fig. 6 and (4375) Kiyomori plots in the basaltic achondrite region (rectangle box). Reddy et al. (2009) briefly discussed the composition of Kiyomori and suggested a surface assemblage dominated by pyroxene, similar to a basaltic achondrite. Using calibrations by Burbine et al. (2009) Kiyomori has a mean pyroxene chemistry of $Fs_{42}En_{49}Wo_9$ and Potapov has $Fs_{29}En_{67}Wo_4$. This suggests that Kiyomori has a pyroxene chemistry that is more Fe-rich than Potapov and it plots in the cumulate eucrite region of the pyroxene quadrilateral plot. In contrast, Potapov's mean pyroxene chemistry suggests a more diogenite-like assemblage. (24245) Ezratty shows two shallow absorption features superimposed on a red slope (increasing reflectance with increasing wavelength). Using equations from Burbine et al. (2009) we have estimated the mean pyroxene chemistry of Ezratty as $Fs_{30-41}En_{50-65}Wo_{5-9}$. This large range in pyroxene chemistry is due to the uncertainties in band II centers and band I centers. We also compared these values with those obtained by more general pyroxene calibration by Gaffey et al. (2002), ($Fs_{34-47}En_{44-63}Wo_{3-9}$) and they agree very well with each other strengthening the case. This chemistry range plots (square symbols connected by solid black line) between the cumulate eucrites and diogenites on the pyroxene quadrilateral (Fig. 9). The band parameters of Ezratty are also consistent with HEDs from Vesta.

The spectrum of (24245) Ezratty (with exception of a shallower Band II) is similar to that of (5404) Uemura, which has been interpreted as a mesosiderite analog (Reddy at al. 2010). Based on the pyroxene chemistry, metal content and spectral similarities, we suggest mesosiderites as the most probable meteorite analogs for Ezratty.

Proper elements of the two S(VII) subtypes (13480) Potapov and (24245) Ezratty and the basaltic achondrite (4375) Kiyomori place them within the Vesta dynamical family. The pyroxene chemistry of (13480) Potapov ($Fs_{29}En_{67}Wo_4$) is within the range for diogenites ($Fs_{20-33}$) and Kiyomori's mean pyroxene chemistry of $Fs_{42}En_{49}Wo_9$ is very similar to cumulate eucrites ($Fs_{32-46}$) or howardites. (24245) Ezratty's pyroxene chemistry ($Fs_{30-41}En_{50-65}Wo_{5-9}$) plots between the cumulate eucrites and diogenites and the domination of metal link it to the mesosiderites meteorites (Mittlefehldt et al. 1998). All three objects could be compositionally related to Vesta based on the mineralogy and pyroxene mineral chemistry. We have also explored the possibility of these three objects being dynamically related to Vesta and Vestoids in the inner Main Belt. It is difficult to dynamically disentangle the Baptistina and Vesta families due to their proximity and the density of objects in this region of the Main Belt.

Figure 10 presents a dynamical map of the proper element space surrounding 298 Baptistina. Included in this figure are Vestoids that have been taxonomically classified as V-types (Bus & Binzel, 2002, Lazzaro et al. 2004, Alvarez-Candal et al. 2006). It is likely that these V-types are genetically related to one another as fragments from the surface of Vesta (Moskovitz et al. 2010). In terms of Δv velocity (Zappala et al. 1996), these V-types extend from the core of the Vesta family to orbits that are up to 2 km/s away from Vesta. Thus, the two S(VII) and BA asteroids, with values of Δv less than 1.5 km/s relative to Vesta, could be interlopers in the BAF. Simulations that account for dynamical evolution due to the Yarkovsky force have shown that Vestoids can diffuse across much of the inner Main Belt and are able to reproduce orbits like those occupied by (13480) Potapov, (24245) Ezratty and (4375) Kiyomori (Nesvorny et al. 2008). Therefore, these three asteroids can plausibly be linked to the Vesta.

**6. Do non-S- and V-type asteroids in BAF have CM2 carbonaceous chondrite composition?**

The mineralogical analysis of (298) Baptistina and the S- and V-type asteroids in its family suggests that they could be part of the Flora and Vesta families respectively. Yet a third group of asteroids in our sample emerged from our study that looked spectrally different from traditional S- and V-types. Five (Fig. 11) of the 16 asteroids in the sample with NIR spectra (Genichesk, Jankovich, 1999 VT23, 1999 WD3 and 1987 SP1) show a clear Band I absorption band but no detectable Band II feature. Initially we suspected these asteroids to be single band asteroids that could be part of the 'real' Baptistina Asteroid Family. Upon closer examination it was determined that the presence of a moderate (<15%) Band II feature can't be ruled out due to low SNR the noise beyond 1.5 μm.

Morphologically, the 1-μm band of Genichesk is similar to phyllosilicate absorption features in carbonaceous chondrites as noted by Reddy et al. (2009). In contrast the Band I absorption feature of (6589) Jankovich, (21910) 1999 VT23, and (12334) 1992 WD3, look similar to silicate features (olivine and low-Fe pyroxene mixture). Reddy et al. (2009) observed (2093) Genichesk and suggested a carbonaceous chondrite (similar to CM2 or CO3) as a possible meteorite analog based on its weak absorption ~1-μm band and spectral similarities. (6589) Jankovich displays a weak absorption feature (band depth ~6.5±0.5%) with a band center of 0.98±0.01 μm on a slight reddish slope. (21910) 1999 VT23 displays a moderately deep (12±0.5%) absorption band at 0.95±0.01 μm on a reddish slope. (12334) 1992 WD3 has a steep slope with a ~1-μm band at 0.96±0.01 μm and a band depth of ~6.3±0.5%.

Unlike Genichesk, which has a shallow, broad bottomed absorption feature (Fig. 11), the other three asteroids have a well-defined Band I absorption band. Finally, the spectrum of (14833) 1987 SP1 shows the deepest 1-μm absorption band (16%) of all the asteroids in this group. Based on the Band I center, band depth, and the presence of a side lobe at 1.3-μm the most dominant mineral in the surface assemblage would be olivine.

The sharp rise in reflectance beyond 2.2µm might be due to poor observing conditions or could be due to a standard star that was not a good solar analog, rather than thermal tail of the Planck curve seen in the spectra of low-albedo NEAs (Reddy, 2009).

Spectral curve matching is the simplest method for comparing spectra of two objects for any affinities in surface composition. Based on silicate-like Band I absorption bands of most 'single band' asteroids we compared their spectra to that of (298) Baptistina. Figure 12 shows an example where spectrum of 1999 VT23 is plotted with Baptistina. Both spectra are normalized to unity at 1.4 microns. Shortward of 1.8 microns both spectra match very well except for the reddish slope. To check for any mineralogical affinity between the 'single band' asteroids and Baptistina we calculated their olivine and pyroxene chemistries (Table 4) using Eq. 2 and 3. Table 4 also shows the olivine and pyroxene chemistries of Baptistina and LL chondrites. Except for slightly lower Fs values of two objects (1992 WD3 and 1999 VT23) the chemistries are consistent with LL-chondrites. This further supports the possibility that a weaker Band II feature could be present but remains obscured by the noise beyond 1.5 microns. The spectral and mineral chemistry match with LL chondrites opens the possibility for these objects to be genetically linked to (298) Baptistina.

Prior to Reddy et al. (2009) and Carvano & Lazzaro (2010) no albedo measurements were available for BAF members. Based on the SDSS colors (C-type), Bottke et al. (2007) assumed a low albedo of 0.04 for the whole BAF. However thermal IR observations of (298) Baptistina have shown that it has a moderate-high albedo (Reddy et al. 2009, and Carvano & Lazzaro, 2010). Albedo of five BAF members were obtained as part of Spitzer MPISGAL Legacy Program (Ryan et al. 2008) using the MIPS instrument (24 µm data) on the Spitzer Space Telescope. The albedos were calculated using Standard Thermal Model (Lebofsky et al. 1986). Table 5 shows the albedo of the observed asteroids along with their SDSS taxonomic classification, when available. Albedos of these asteroids range from moderate (0.14) to high (0.35), which would be inconsistent with a carbonaceous chondrite or C-type interpretation. Albedos of CM2 carbonaceous chondrites range from 0.03-0.06 (Gaffey, 1976) and from 0.22-0.37 for LL chondrites (Gaffey, 1976) with an average 0.27. The average albedo of BAF based on five measurements from Spitzer and Baptistina (0.347) from Carvano & Lazzaro (2010) is 0.26, which would be consistent with average albedo of LL-chondrites. It is clear from the mineralogy and albedo measurements that even smaller non-S and V-type ('single' band) asteroids in Baptistina family do not have surface composition similar to CM2 carbonaceous chondrites.

While the dominant mineralogy of the asteroids in the sample appears to be similar to LL chondrites, an important spectral difference can be noted between their spectra, namely absorption band depth. As already noted Baptistina's Band I and II depths are shallower than normal for S-type asteroids, and two other 'single band' asteroids in our sample (Jankovich and 1992 WD3) also show weaker bands (Band I depth 6-7%). Reddy et al. (2009) noted several factors (abundance of absorbing species, metal, carbon, phase angle, particle size, packing) that could affect the absorption band depth.

# 7. Relations among SDSS colors, visible and NIR spectra and geometric albedos

The number of asteroids in the Baptistina region with observations in the SDSS

MOC4 surpasses, by several orders of magnitude, that of BAF members with NIR data. It is thus useful to examine, whenever possible, the correspondence between the mineralogical inferences made from NIR spectra to visible spectra and SDSS colors. Table 6 lists the family members, according to the adopted family definition, with either a visible spectrum or mineralogical interpretation through NIR spectrum, showing for each asteroid the mineralogical group (see Section 3), the taxonomic class (Bus and Binzel 2002) of the visible spectra, and the taxonomic classification from SDSS colors and respective likelihood of the classification (Carvano et al. 2010). Figure 9 shows the visible spectrum, with SDSS color, whenever available.

There are 6 asteroids with both NIR and visible spectra: two Xc-, one C-, one S-, and two A-type asteroids. Both Xc-type asteroids, (298) Baptistina and (12334) 1992 WD3, present NIR spectra with suppressed absorption bands; for (298) Baptistina this probably corresponds to an LL-chondrite mineralogy, while (12334) 1992 WD3 possibly has a pyroxene-dominated mineralogy. The two Bus A-type asteroids, on the other hand, show rather diverse mineralogies: (1126) Otero has a mineralogy compatible with LL-chondrites, while (4375) Kiyomori is similar to basaltic achondrites, but with considerable suppression of the 1 and 2 micron bands. The single C-type asteroid, (2093) Genichesk, has a NIR spectrum that is similar to some carbonaceous chondrites. Finally, the S-type asteroid (1365) Henyey has mineralogy, which is also compatible with LL-chondrites and very similar to (298) Baptistina and (1126) Otero.

There are also 6 asteroids with both NIR spectrum and SDSS colors: three $S_p$-, one $CX_p$-, one $Q_p$- and one $QO_p$-type. In the taxonomy based on SDSS photometry (hence the subscript "p"), derived by Carvano et al (2010), the likelihood of each classification is given as an integer from 0 to 100. Two of the $S_p$-type asteroids, (3260) Vizbor and (6266) Letzel, have classification with likelihood greater than 50 and present a S(III)-type NIR spectra. The remaining $S_p$ asteroid, (14833) 1987 SP1, has rather low classification likelihood and has a NIR with a somewhat suppressed 1-μm band. The $CX_p$-type asteroid (12334) 1992 WD3 has a classification likelihood of 49 and presents a NIR spectrum corresponding possibly to a pyroxene-dominated mineralogy with a suppressed 1-μm band and no visible 2-μm band. This type of NIR spectra is also seen on the $Q_p$- and $O_p$-type asteroids, (24245) Ezratty and (21910) 1999 VT23, whose taxonomic classification have otherwise rather low likelihoods. Finally, only three asteroids have both visible spectra and SDSS colors – (7255) 1993 VY1 ($C_p$), (12334) 1992 WD3 ($CX_p$), and (7479) 1994 EC1 ($S_p$) – and for all three the classifications made from SDSS photometry are reasonably compatible with the spectroscopic classification.

Therefore, the C and X taxonomic types in the BAF region -- including (298) Baptistina -- are more frequently associated with NIR spectra that display a suppressed 1-μm band and seem to have a mineralogy dominated by pyroxene mixed with a spectrally neutral phase that suppresses the bands. This is supported by the albedos of some C-, X-type family members that have intermediate to high values (0.14-0.33), unlike carbonaceous chondrites. Finally, the S-type asteroids in the BAF region seem to have mineralogies compatible with the S(IV) and S(III) subtypes.

**7.1 Distribution of Taxonomies and Mineralogies in the Family**

Figures 13-15 show the density distribution (background color) in proper element space of objects with SDSS taxonomic classification, using bins of 0.008 AU in proper

semi major axis and 0.4 degrees in proper inclination for the C- and X-, S-, and V-type asteroids respectively. The location of the asteroids in our sample with surface mineralogies derived from NIR data are marked in these figures. The black color symbols represent asteroids with suppressed absorption bands.

The BAF is seen in Fig. 13 as a two-lobbed density enhancement of C- and X-type asteroids, with the brightest (and presumably bigger) objects concentrated in the relatively low-density region between the two lobes. Such configuration is believed to be produced by the Yarkovsky drift of the family members (Vokrouhlický et al. 2006) and is seen in other asteroid families. The lower density of the inner lobe as well as the spread in proper inclination may be explained by the dynamics of the region. There are some interesting facts to be gathered from the position of the asteroids with NIR spectra in Fig. 13: i) The asteroids with suppressed band in their spectra tend to be concentrated in the central void and in the higher density outer lobe; ii) the S(IV) asteroids are either in the central void or close to higher density clumps in the inner lobe; iii) (9143) Burkhead and one of the S(III) asteroids are located far from the higher density regions. It should be noted however that the number of objects with NIR spectra is relatively small compared to the total number of objects in the family.

In contrast, no clear density enhancement is seen in the distribution of S-type asteroids that could be clearly associated with a family (Fig. 14). Although there is an enhancement in the density of S-type asteroids centered on Baptistina, it is too dispersed in inclination compared to other asteroid families. For instance, the high-density stripe starting at 2.32 AU and around proper inclination of 3.9º is part of the Massalia family.

Finally, Fig. 15 shows the density distribution for the V-type asteroids. The high density region starting at around 2.5 AU and at proper inclination of around 7 degrees is the Vesta family. Comparing with Fig. 13 it is clear that the outer lobe of the Baptistina family, although located at lower inclinations, has some overlap in semi major axis with the Vesta family. It is also interesting to note that the basaltic asteroid with suppressed bands, (4375) Kiyomori, is located closer to the Vesta family than (13480) Potapov, which has a NIR spectra similar to what is expected from a typical Vestoid. (4375) Kiyomori is also located very near the highest density region of the outer lobe in the distribution of C- and X-type asteroids (Fig. 13).

It is instructive to consider how the family membership of the objects with NIR spectra depends on the adopted cut-off. At 45 m/s only (298) Baptistina and (2093) Genichesk are in the family, which includes only the densest part of the outer lobe. As the cut-off increases to 50 m/s the family includes most of the outer, densest lobe, but only 6 of the objects with NIR spectra are formal members at this level (298, 2093, 4375, 6589, 12334 and 21910), and all of them present suppressed features or are featureless. Finally, at a cut-off level of 56 m/s the objects at the inner, less dense lobe are included into the family. However, 5 of the objects with observed NIR spectra are still not part of the family at this level: the D-type (9143), the V-type basaltic asteroid (13489), two S(III) asteroids (6266 and 13154), and the S(IV) asteroid (1126) that plots closer to (298) Baptistina in a vs. I, but which is outside the range of the proper eccentricities of the family members. Again, this suggests that the Baptistina family is defined by objects with suppressed olivine/pyroxene bands (which appear as C-, X-type in SDSS colors and visible spectra) and that the asteroids in the region with more normal Ol/Px bands (S) are either background objects or were more efficiently dispersed than their C-, X-type counterparts.

## 7. Conclusions

Our mineralogical study of the Baptistina asteroid family has answered several key questions about the nature, origin and its relationship with the K/T impactor source. Here we summarize key findings of our study.

1) The Baptistina family lies in a dynamically complex region and can only be robustly defined using SDSS colors, since the family is defined by a concentration of asteroids with C/X-type colors in a region whose background is dominated by S-type colors.

2) The brightest asteroid in the family, (298) Baptistina, is compatible with the other asteroids are similar in terms of colors, albedo, and silicate composition and, therefore, there is no reason to consider it as an interloper.

3) (298) Baptistina is neither a C-type nor does it have a surface composition analogous to CM2 carbonaceous chondrite, but is similar to LL chondrites based on mineralogy and albedo.

4) The smaller BAF asteroids with lower SNR spectra showing only a 'single', suppressed band are compositionally similar to (298) Baptistina and L/LL chondrites.

5) The S-type and V-type asteroids we characterized in and around the vicinity of BAF show mineralogical affinity to (8) Flora and (4) Vesta and could be part of their families.

The simplest scenario is, therefore, to assume that the Baptistina family was formed from the breakup of a parent body with an LL-chondrite composition. However, it is not clear at this point why the silicate bands observed in the asteroids with formal family definition seem suppressed with respect to the background population. This is particularly puzzling given that, in terms of silicate mineralogy, the composition of the family is similar both to the background population and to the nearby Flora family. At any rate, the silicate composition of the family and its albedo are in disagreement with the hypothesis used in Bottke et al. (2007) to conclude that the Baptistina family was the source of the K/T impactor. Since the core family shows a high degree of mineralogical homogeneity, consistent with a LL-chondrite composition, this family cannot be the source of the K/T impactor, if indeed that body had a CM2-like composition as implied by analysis of sediments and meteorite fragments from the K/T boundary (REFs 40, 41 and 42 in B2007). Moreover, the albedo of the family turns out to be much higher than the value of 0.04 used by Bottke et al (2007) to date the formation of the family. Without changing any other parameters in the method used by Bottke at al. (2007) to date the family, such higher albedo would yield a much younger age to the Baptistina family. Therefore, whatever the K/T impactor might have been, there is no reason to assume at this point that it is related to Baptistina.

**Acknowledgements**

V.R., M.J.G. were supported by NASA NEOO Program Grant NNX07AL29G and NASA Planetary Geology and Geophysics Grant NNX07AP73G. V.R. was also supported by a PCI/MCT fellowship during his stay at Observatório Nacional. J.M.C., D.L. and T.M.D. were supported by several fellowship and grants by the Brazilian National Research Council – CNPq and the Rio de Janeiro Science Foundation – FAPERJ. T.A.M. work has been supported by the Brazilian National Research Council -

CNPq, the São Paulo State Science Foundation – FAPESP; she gratefully acknowledges the support of the Computation Center of the University of São Paulo (LCCA-USP) and of the Astronomy Department of the IAG/USP, for the use of their facilities. The research utilizes spectra acquired by C. M. Pieters, and P. Hudson with the NASA RELAB facility at Brown University. The authors would like to thank Bill Bottke and Tasha Dunn for their reviews and comments to improve the manuscript; Martin Hynes, Bobby Bus, M. Schaal, and Driss Takir for their support in this research. We thank the IRTF TAC for awarding time to this project, and to the IRTF TOs and MKSS staff for their support.

**Figure Captions**

Figure 1. Location of the Baptistina family (cyan) in dynamical space. The random black points are asteroids and the dotted curves are secular resonances. The magenta symbol is the location of asteroid (8) Flora and the red symbol is the location of (298) Baptistina. The green family located at proper inclination 1.4° is Massalia family and the yellow family at located at proper inclination 2.6° is the Nysa family.

Figure 2. Number of possible Baptistina family members as a function of escape velocity cutoff.

Figure 3. Clusters of objects (cyan) in the Baptistina region for different values of escape velocity cutoff. The random black points are asteroids and the dotted curves are secular resonances. The magenta symbol is the location of asteroid (8) Flora and the red symbol is the location of (298) Baptistina.

Figure 4. Baptistina family (cyan) members detected by applying HCM to C-/X-class sample.

Figure 5. Visible (black) and NIR spectra (red) of asteroids belonging to the Gaffey S(IV) subtype of S-asteroids from the Baptistina Asteroid Family. Visible data was acquired from 1.5-m telescope at the ESO, La Silla, Chile, and the 3.58 m TNG, La Palma, Canary Islands, and the NIR data was acquired at the NASA IRTF, Mauna Kea, Hawai'i.

Figure 6. Band I center vs. Band Area Ratio plot showing the various Gaffey S-asteroid subtypes. Spectral band parameters of BAF members with two absorption bands are plotted to show their affinity to a specific subtype.

Figure 7. Broadband colors (green) and NIR spectra (red) of Gaffey S(III) subtypes from the Baptistina Asteroid Family.

Figure 8. Broadband colors (green), visible (black) and NIR spectra (red) of Gaffey (VII)/Basaltic subtypes from the Baptistina Asteroid Family. Board band colors are from Sloan Digitized Sky Survey, visible data was acquired from 1.5-m telescope at the ESO, La Silla, Chile, and the 3.58 m TNG, La Palma, Canary Islands, and the NIR data was acquired at the NASA IRTF, Mauna Kea, Hawai'i.

Figure 9. Pyroxene quadrilateral plot showing enstatite (En), ferrosilite (Fs), Diopside (Di) and hedenbergite (Hd) end members. The pyroxene chemistry zones (from left to right) for diogenites (left oval), cumulate eucrites/howardites (center rectangle), and basaltic eucrites (right oval). Mean pyroxene chemistry of the three S(VII)/basaltic asteroids is plotted (from left to right) (13480) Potapov (triangle), (24245) Ezratty (square), and (4375) Kiyomori (diamond). Due to Band center uncertainties Ezratty's pyroxene chemistry range is plotted (squares connected with solid line).

Figure 10. Dynamical map of Vesta family showing the location of S(VII) (star) and basaltic achondrite (inverted triangle) members of the BAF along with Main Belt

asteroids (4) Vesta, (8) Flora and (298) Baptistina. Delta V values for the three BAF members are also shown. The open triangles are V-type asteroids.

Figure 11. Broadband colors (green), visible (black) and NIR spectra (red) of BAF asteroids that display only a single absorption band. Board band colors are from Sloan Digitized Sky Survey, visible data was acquired from 1.5-m telescope at the ESO, La Silla, Chile, and the 3.58 m TNG, La Palma, Canary Islands, and the NIR data was acquired at the NASA IRTF, Mauna Kea, Hawai'i. Note the high scatter beyond 1.4 microns is due to large atmospheric residuals which

Figure 12. Spectrum of 1999 VT23 (Red) is plotted with (298) Baptistina (black). Both spectra are normalized to unity at 1.4 microns.

Figure 13. Density distribution of C-class objects near Baptistina Asteroid Family in proper element space (inclination vs. semi-major axis). BAF members are shown here along with their Gaffey S-asteroid subtype affinities and dominant mineralogy (BA - basaltic achondrites; PY - pyroxene; Ol - olivine), meteorite affinity (CC - carbonaceous chondrite) or taxonomy (D-type).

Figure 14. Density distribution of S-class objects near Baptistina Asteroid Family in proper element space. The high-density stripe (black oval) starts at 2.32 AU and around proper inclination of 3.9º is part of the Massalia family. The S(III) subtypes are filled squares, S(IV) subtypes (open triangles), S(VII) subtypes (filled star), BA-basaltic

achondrites (filled triangle), Px-pyroxene dominated (filled diamond), Ol-olivine dominated (open star), CC-carbonaceous chondrite (open diamond), and D-type asteroid (open cross). The filled symbols represent objects with suppressed bands.

Figure 15. Density distribution of V-class objects near Baptistina Asteroid Family in proper element space. The symbols used in this figure are similar to those in Figure 13.

**Table 1.** Observational circumstances for asteroids with visible spectra

| Asteroid | Date | Mag. V | Airmass | Phase Angle | Helio. Dist. |
|---|---|---|---|---|---|
| (298) Baptistina | 4-Jan-97 | 14.4 | 1.86 | 1.86 | 14.4 |
|  | 15-Dec-06 | 14.9 | 1.04 | 1.04 | 14.9 |
| (525) Adelaide | 4-Jan-97 | 15.5 | 1.21 | 1.21 | 15.5 |
| (711) Marmulla | 2-Sep-02 | 14.5 | 1.00 | 1.00 | 14.5 |
| (2093) Genichesk | 6-Jan-97 | 16.8 | 1.39 | 1.39 | 16.8 |
|  | 28-Oct-06 | 15.9 | 1.07 | 1.07 | 15.9 |
| (2472) Bradman | 16-Dec-06 | 16.4 | 1.01 | 1.01 | 16.4 |
| (2768) Gorky | 3-Sep-02 | 14.2 | 1.03 | 1.03 | 14.2 |
| (2858) Carlosporter | 1-Sep-02 | 15.7 | 1.14 | 1.14 | 15.7 |
| (2961) Katsurahama | 2-Jan-97 | 15.7 | 1.29 | 1.29 | 15.7 |
| (3533) Toyota | 4-Jan-97 | 15.8 | 1.47 | 1.47 | 15.8 |
| (4263) Abashiri | 2-Sep-02 | 15.7 | 1.03 | 1.03 | 15.7 |
| (4278) Harvey | 8-Jan-97 | 15.8 | 1.38 | 1.38 | 15.8 |
|  | 28-Oct-06 | 15.8 | 1.16 | 1.16 | 15.8 |
| (4375 Kiyomori | 21-Nov-96 | 16.3 | 1.63 | 1.63 | 16.3 |
| (5238) Naozane | 21-Mar-02 | 15.1 | 1.16 | 1.16 | 15.1 |
| (6253) 1992FJ | 25-Mar-02 | 16.9 | 1.24 | 1.24 | 16.9 |
| (7255) 1993 VY1 | 20-Mar-02 | 16.2 | 1.13 | 1.13 | 16.2 |
| (7479) 1994 EC1 | 28-Oct-06 | 16.5 | 1.05 | 1.05 | 16.5 |
|  | 15-Dec-06 | 16.7 | 1.08 | 1.08 | 16.7 |
| (9061) 1992 WC3 | 16-Dec-06 | 17.2 | 1.01 | 1.01 | 17.2 |
| (11913) Svarna | 15-Dec-06 | 16.9 | 1.01 | 1.01 | 16.9 |

| (12334) 1992 WD3 | 16-Dec-06 | 17.4 | 1.07 | 1.07 | 17.4 |
| (44063) 1998 FW50 | 28-Oct-06 | 17.9 | 1.03 | 1.03 | 17.9 |

**Table 2.** Observational circumstances for asteroids with near-IR spectra

| Asteroid | Date | Mag. V | Airmass | Phase Angle | Helio. Dist. |
|---|---|---|---|---|---|
| (298) Baptistina | 21-Mar-08 | 13.9 | 1.25 | 13.9 | 2.13 |
| (1126) Otero | 21-Mar-08 | 14.7 | 1.17 | 18.3 | 1.96 |
| (1365) Henyey | 23-Mar-08 | 14.2 | 1.18 | 8.9 | 2.10 |
| (2093) Genichesk | 23-Mar-08 | 16.4 | 1.04 | 16.9 | 2.39 |
| (4375) Kiyomori | 23-Mar-08 | 14.9 | 1.08 | 5.9 | 2.08 |
| (13154) Petermrva | 1-Sep-08 | 15.6 | 1.18 | 1.5 | 1.94 |
| (13480) Potapov | 1-Sep-08 | 15.9 | 1.28 | 5.9 | 1.74 |
| (2545) Verbiest | 1-Sep-08 | 15.8 | 1.16 | 14.5 | 2.12 |
| (9143) Burkhead | 31-Oct-08 | 15.5 | 1.01 | 5.1 | 1.96 |
| (14833) 1987 SP1 | 31-Oct-08 | 15.9 | 1.07 | 3.9 | 2.00 |
| (3260) Vizbor | 31-Oct-08 | 16.1 | 1.04 | 20.7 | 2.18 |
| (21910) 1999 VT23 | 14-Sep-09 | 15.2 | 1.38 | 3.2 | 1.89 |
| (6589) Jankovich | 20-Oct-09 | 16.1 | 1.02 | 6.2 | 1.95 |
| (12334) 1992 WD3 | 18-Nov-09 | 15.7 | 1.11 | 6.4 | 1.96 |
| (24245) Ezratty | 18-Nov-09 | 16.3 | 1.23 | 2.4 | 1.99 |
| (19739) 2000 AL104 | 20-Oct-09 | 16.5 | 1.05 | 5.1 | 1.27 |
| (6266) Letzel | 21-Jun-06 | 15.9 | 1.10 | 18.1 | 1.94 |

**Table 3.** Spectral band parameters of asteroids with two absorption bands (S-,V-type)

| Object Units > | Band I Center μm | Errors μm | Band I Depth % | Band II Center μm | Errors μm | Band II Depth % | Band Area Ratio | Errors | Olivine % | Pyroxene % | Fa | Fs | Gaffey S-subtype |
|---|---|---|---|---|---|---|---|---|---|---|---|---|---|
| (6266) Letzel | 0.98 | 0.01 | 15.00 | 2.00 | 0.03 | 9.40 | 0.64 | 0.05 | 0.62<br>0.68 | 0.38<br>0.32 | 27 | 22<br>42 | S(III) |
| (3260) Vizbor | 0.99 | 0.01 | 17.00 | 1.97 | 0.03 | 7.00 | 0.58 | 0.07 | 0.64<br>0.70 | 0.36<br>0.30 | 28 | 23<br>24 | S(III) |
| (13154) Petermrva | 1.00 | 0.01 | 17.00 | 1.87 | 0.02 | 10.00 | 0.76 | 0.07 | 0.60<br>0.63 | 0.40<br>0.32 | 29 | 24<br>18 | S(III) |
| (298) Baptistina | 1.00 | 0.01 | 7.00 | 1.96 | 0.02 | 2.00 | 0.31 | 0.05 | 0.70<br>0.82 | 0.30<br>0.32 | 29 | 24<br>21 | S(IV) |
| (1126) Otero | 0.98 | 0.01 | 16.00 | 2.00 | 0.01 | 6.00 | 0.40 | 0.05 | 0.68<br>0.78 | 0.32<br>0.32 | 27 | 22<br>40 | S(IV) |
| (1365) Henyey | 1.01 | 0.01 | 15.00 | 2.01 | 0.01 | 4.50 | 0.29 | 0.05 | 0.70<br>0.83 | 0.30<br>0.32 | 30 | 25<br>42 | S(IV) |
| (2545) Verbiest | 0.96 | 0.01 | 13.00 | 1.90 | 0.03 | 8.00 | 0.45 | 0.07 | 0.66<br>0.76 | 0.34<br>0.32 | 23 | 19<br>26 | S(IV) |
| (13480) Potapov | 0.92 | 0.01 | 51.00 | 1.92 | 0.02 | 39.00 | 1.63 | 0.07 | 0.27 | 0.73 | | 29<br>31 | S(VII) |
| (24245) Ezratty | 0.91 | 0.02 | 4.00 | 1.98 | 0.05 | 5.00 | 2.38 | 0.15 | 0.00 | 1.00 | | 30-41<br>34-47 | S(VII) |
| (4375) Kiyomori | 0.93 | 0.01 | 13.00 | 2.00 | 0.02 | 11.00 | 1.63 | 0.05 | 0.27 | 0.73 | | 42<br>42 | BA |

**Table 4.** Olivine and Pyroxene chemistries of asteroids with single band

| Asteroid/Meteorite | Olivine Fa % | Pyroxene Fs % |
|---|---|---|
| (298) Baptistina | Fa29 | Fs24 |
| (6589) Jankovich | Fa27 | Fs22 |
| (12334) 1992 WD3 | Fa24 | Fs20 |
| (14833) 1987 SP1 | Fa30 | Fs25 |
| (21910) 1999 VT23 | Fa24 | Fs20 |
| LL chondrites | Fa27-33 | Fs23-27 |

**Table 5.** Albedo, diameter and SDSS taxonomic classification of BAF members.

| Object | Albedo | Diameter (km) | SDSS class |
|---|---|---|---|
| 1997 VZ4 | 0.21 | 4.04 | X |
| 1997 WM24 | 0.32 | 2.69 | - |
| 1990 TW8 | 0.22 | 3.24 | C |
| 1999 GN25 | 0.33 | 1.68 | C |
| 2000 NV13 | 0.14 | 2.44 | X |

**Table 6.** Taxonomic classification of BAF members based on SDSS, visible and NIR data.

| Asteroid # | NIR | Vis. Spec. | SDSS |
|---|---|---|---|
| 298 | S(IV) | Xc | - |
| 711 | - | A | - |
| 1126 | S(IV) | A | - |
| 1365 | S(IV) | S | - |
| 2093 | CC(??) | C | - |
| 2472 | - | S | - |
| 2545 | S(IV) | - | - |
| 2768 | - | L | - |
| 2858 | - | L | - |
| 2961 | - | S | - |
| 3260 | S(III) | - | $S_p$ (92) |
| 3533 | - | X | - |
| 4263 | - | S | - |
| 4278 | - | V | - |
| 4375 | BA | A | - |
| 5238 | - | S | - |
| 6253 | - | X | - |
| 6266 | S(III) | - | $S_p$ (51) |
| 6589 | Px(?) | - | - |
| 7255 | - | C | $C_p$ (71) |
| 7479 | - | S | $S_p$ (80) |
| 9061 | - | Xc | - |
| 9143 | D | | - |
| 11913 | - | O | - |
| 12334 | Px(?) | Xc | $CX_p$ (49) |
| 13154 | S(III) | - | - |
| 13480 | BA | - | - |
| 14833 | Ol(?) | - | $S_p$ (11) |
| 19739 | ? | - | $C_p$ (67) |

| 24245 | S(VII) | - | $Q_p$ (10) |
| 21910 | Px(?) | - | $QO_p$ (24) |
| 44063 | - | C | - |

**Figure 1: Baptistina Asteroid Family**

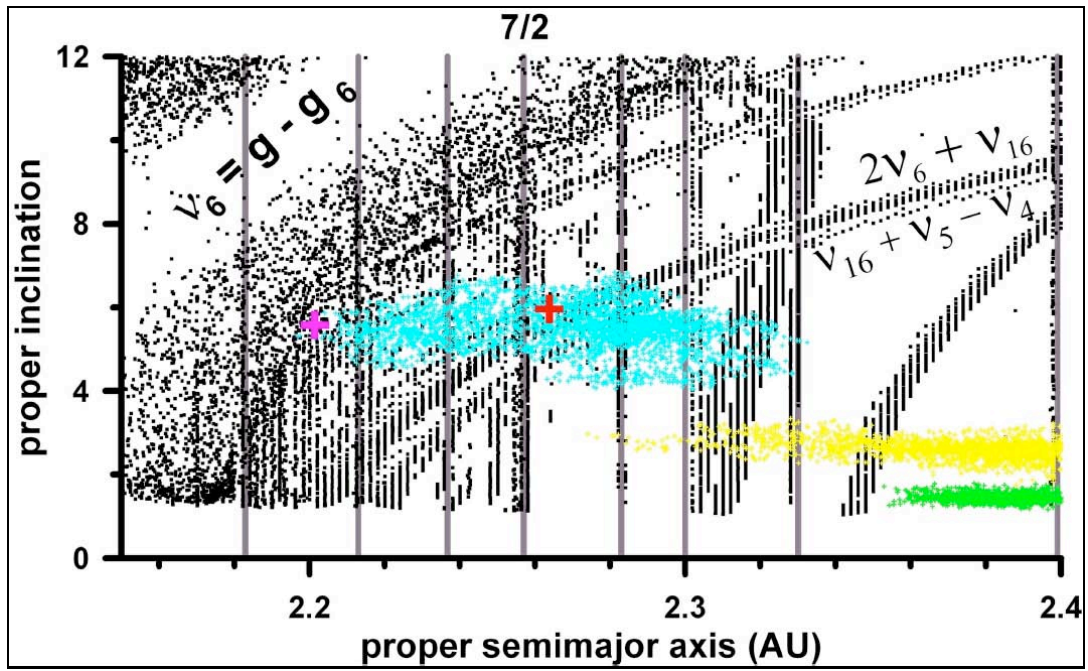

**Figure 2: Baptistina Asteroid Family**

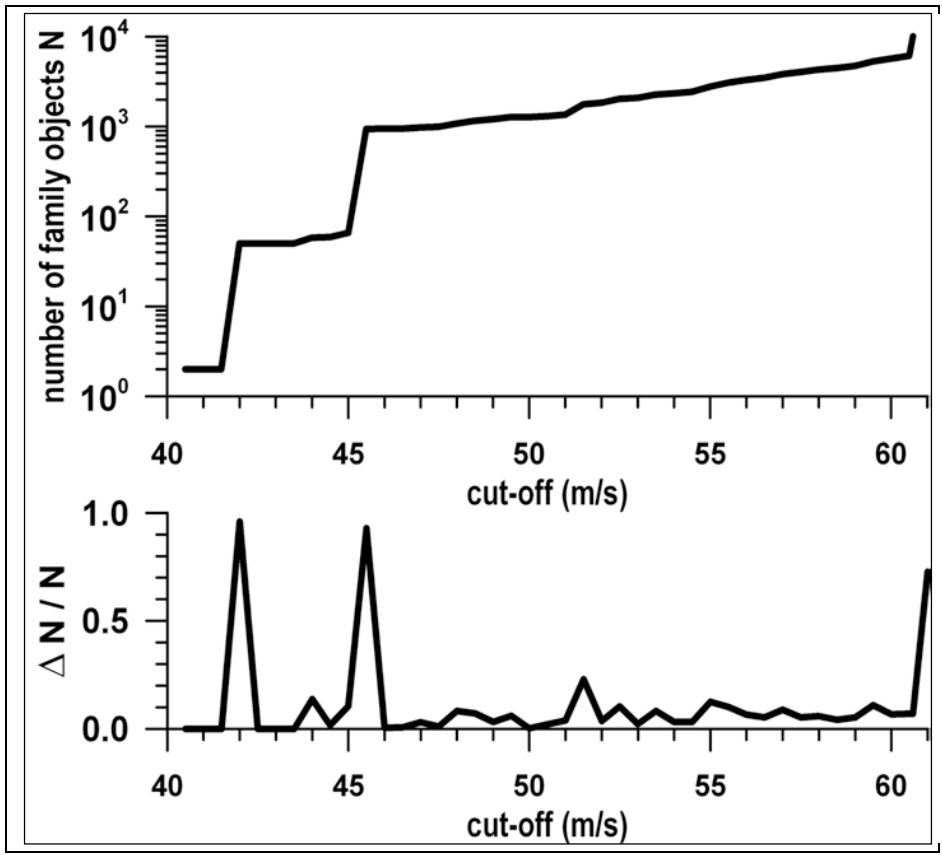

**Figure 3: Baptistina Asteroid Family**

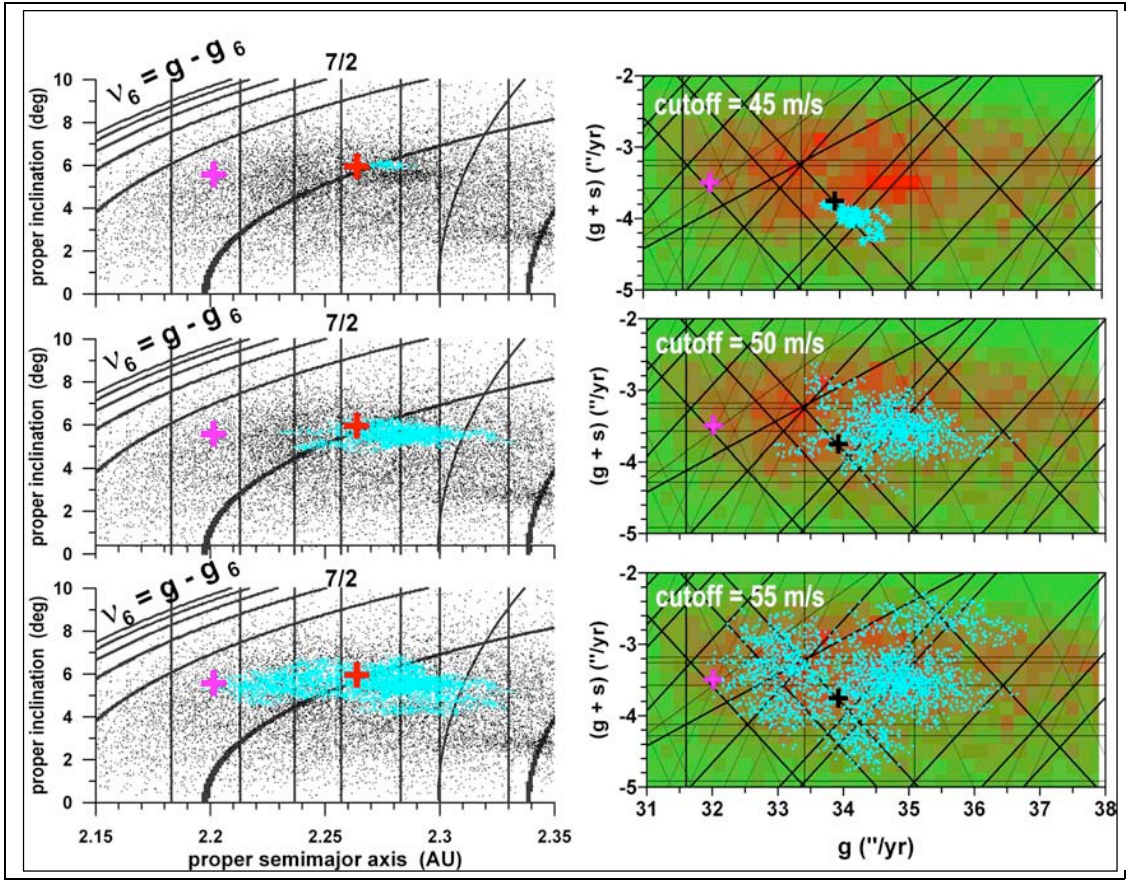

**Figure 4: Baptistina Asteroid Family**

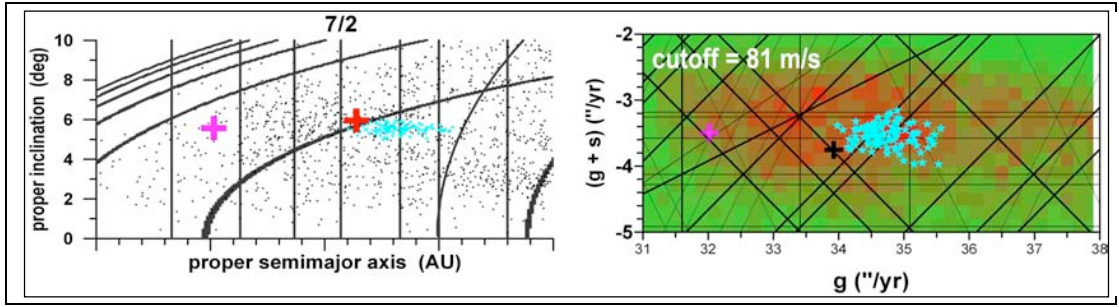

**Figure 5: Baptistina Asteroid Family**

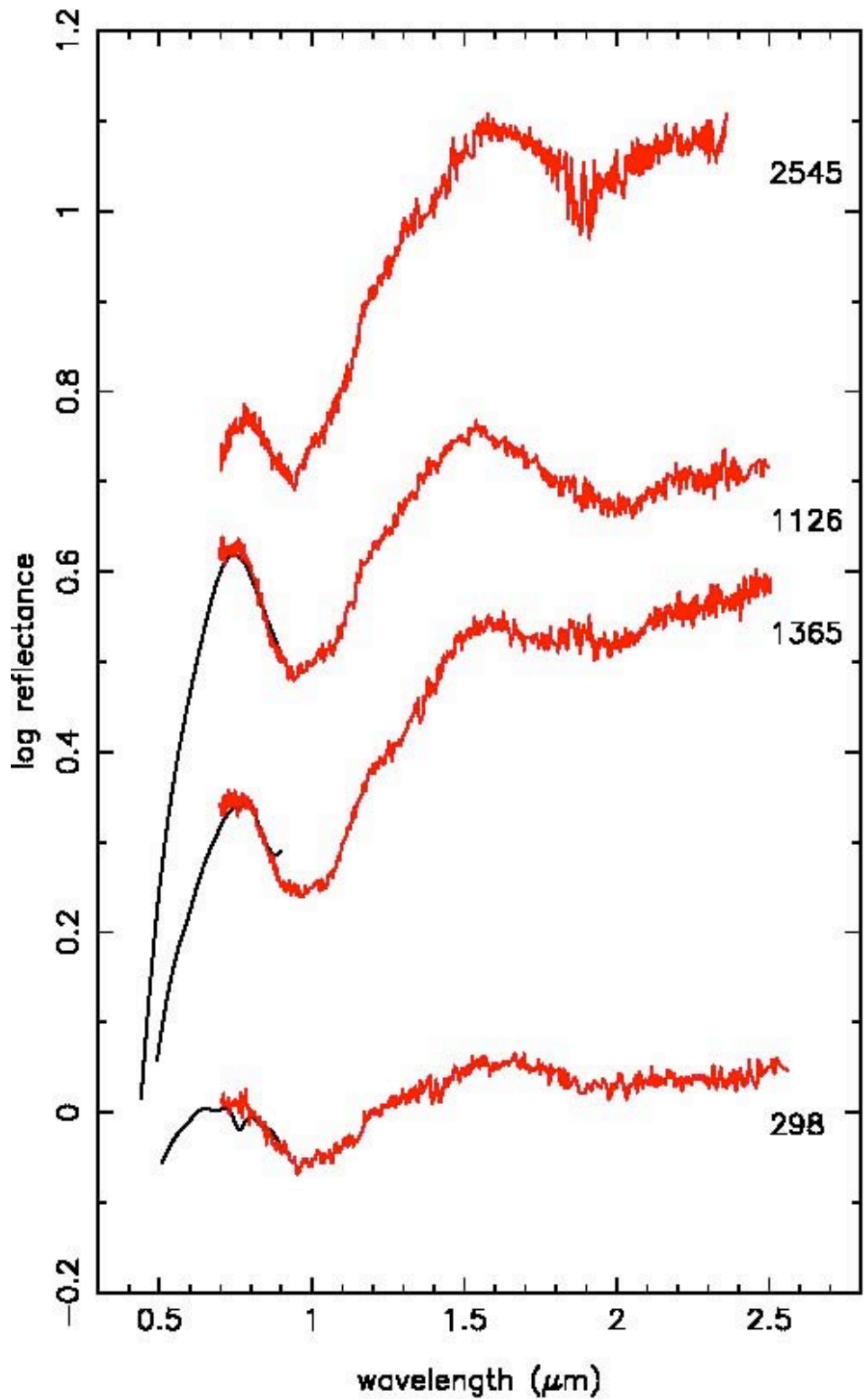

**Figure 6: Baptistina Asteroid Family**

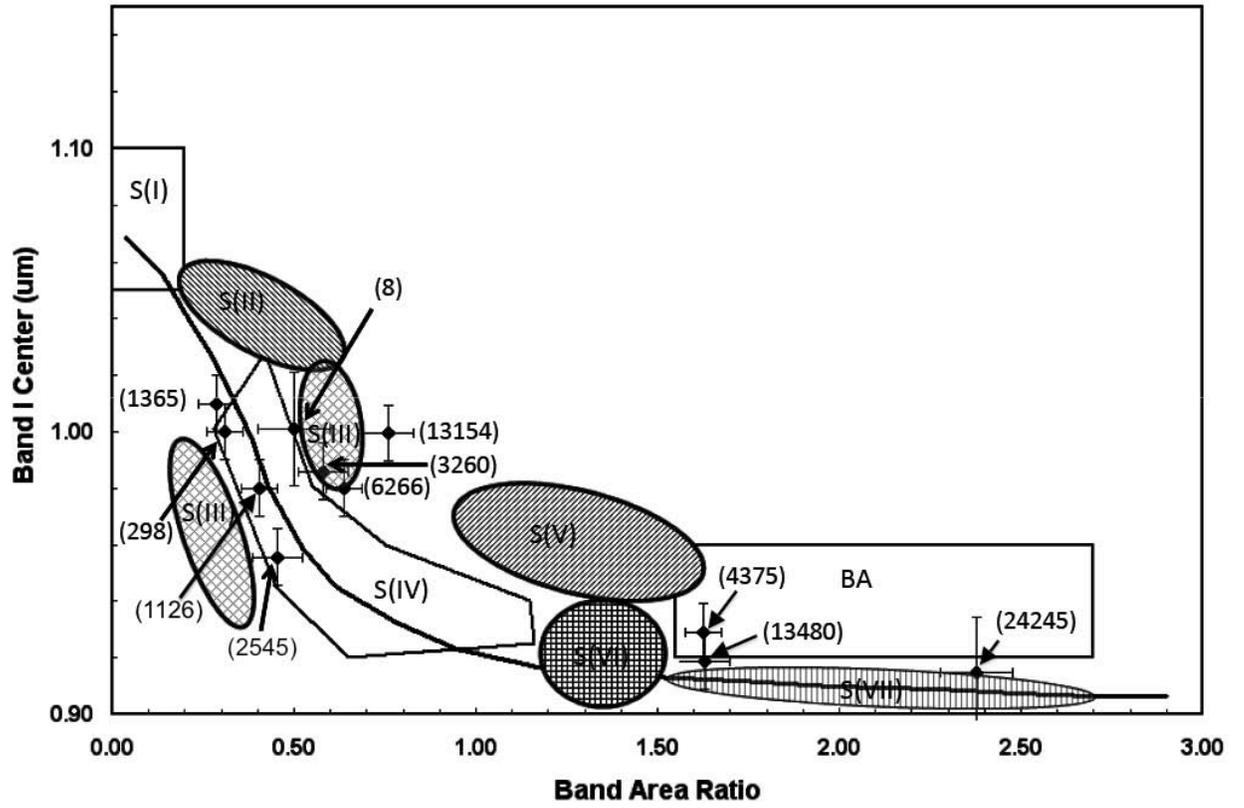

**Figure 7: Baptistina Asteroid Family**

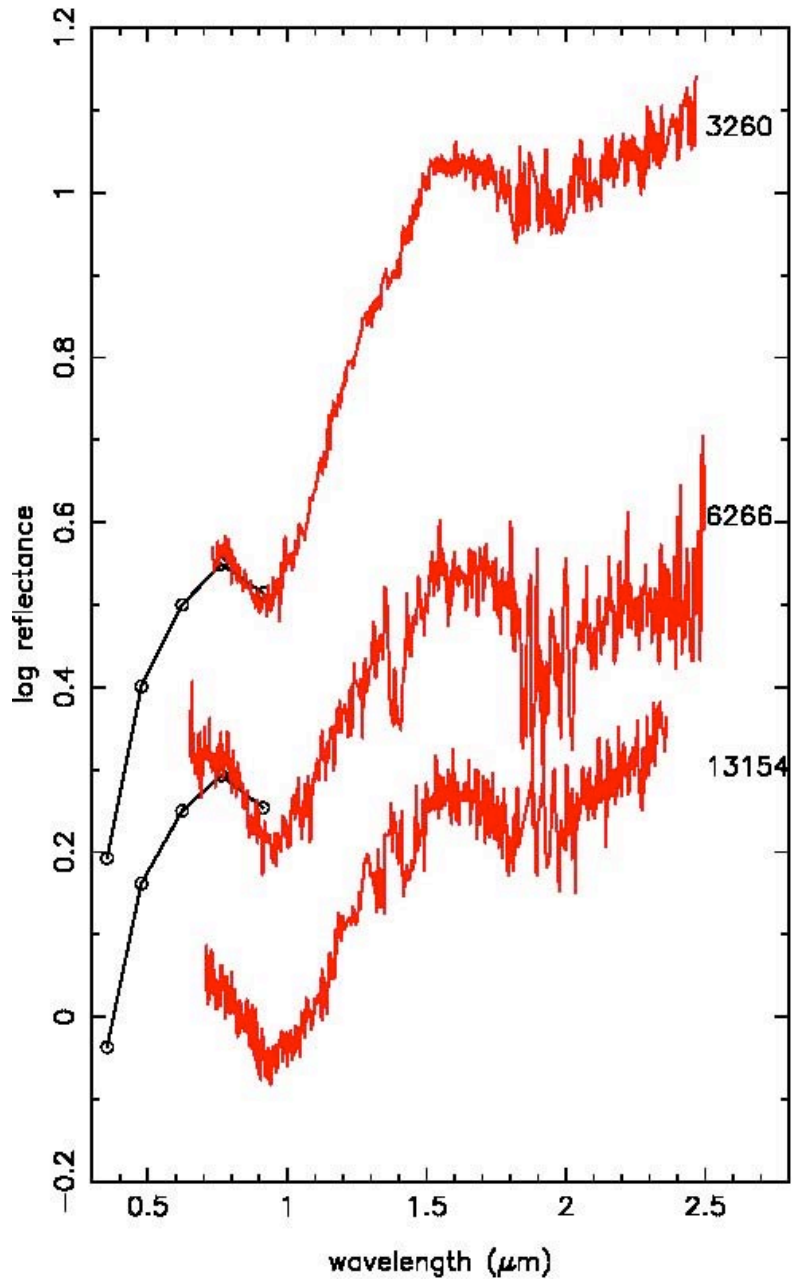

**Figure 8: Baptistina Asteroid Family**

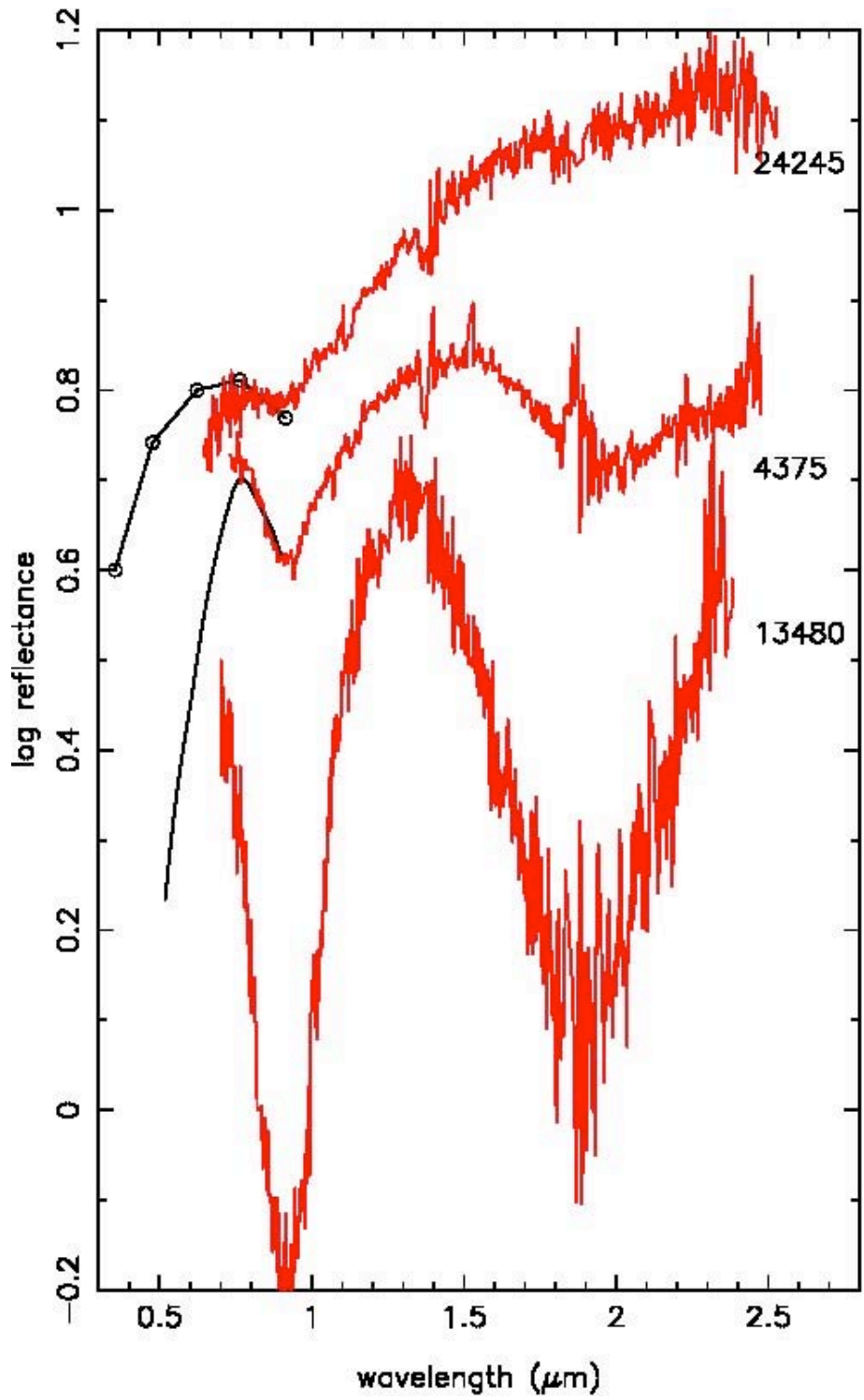

**Figure 9: Baptistina Asteroid Family**

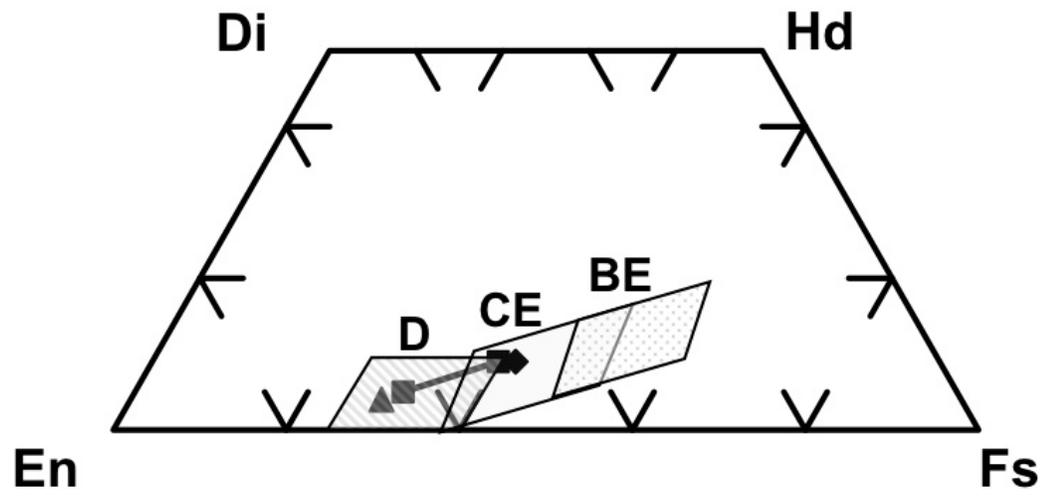

**Figure 10: Baptistina Asteroid Family**

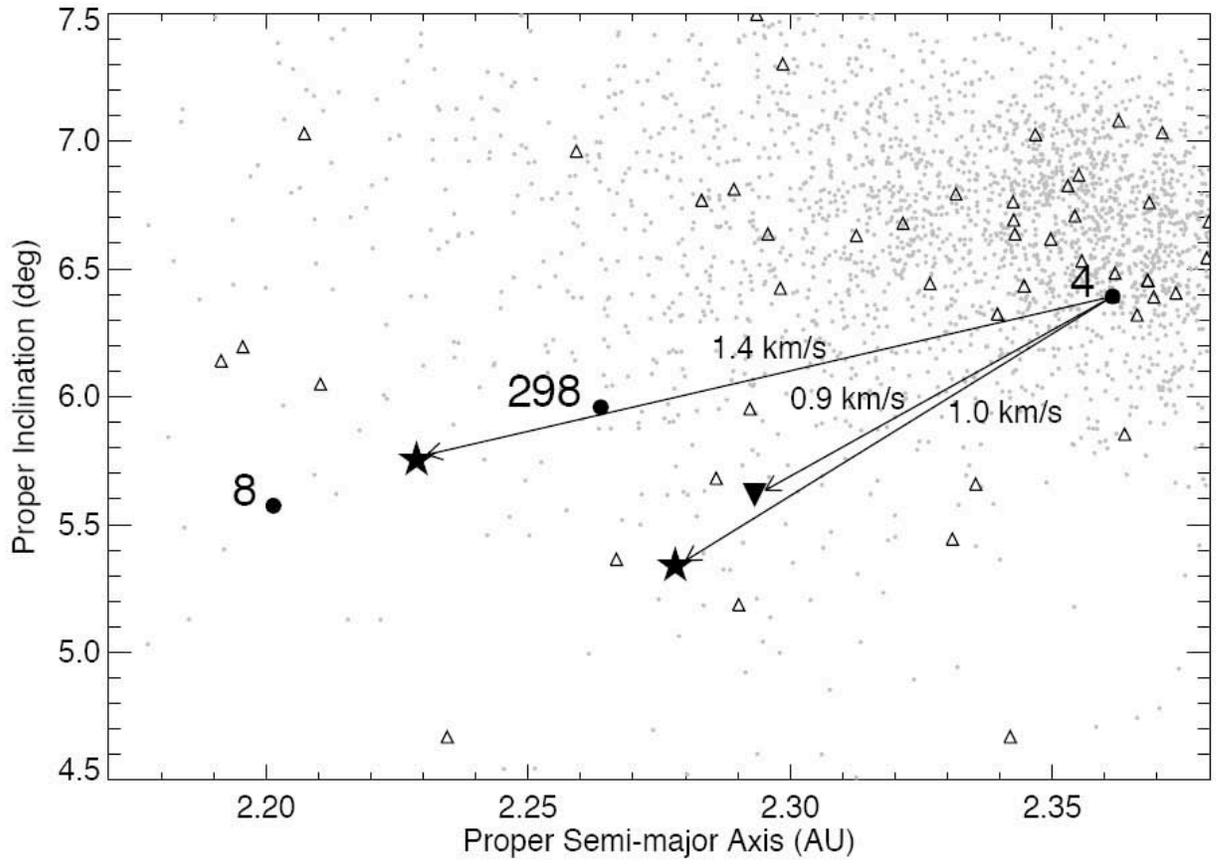

**Figure 11: Baptistina Asteroid Family**

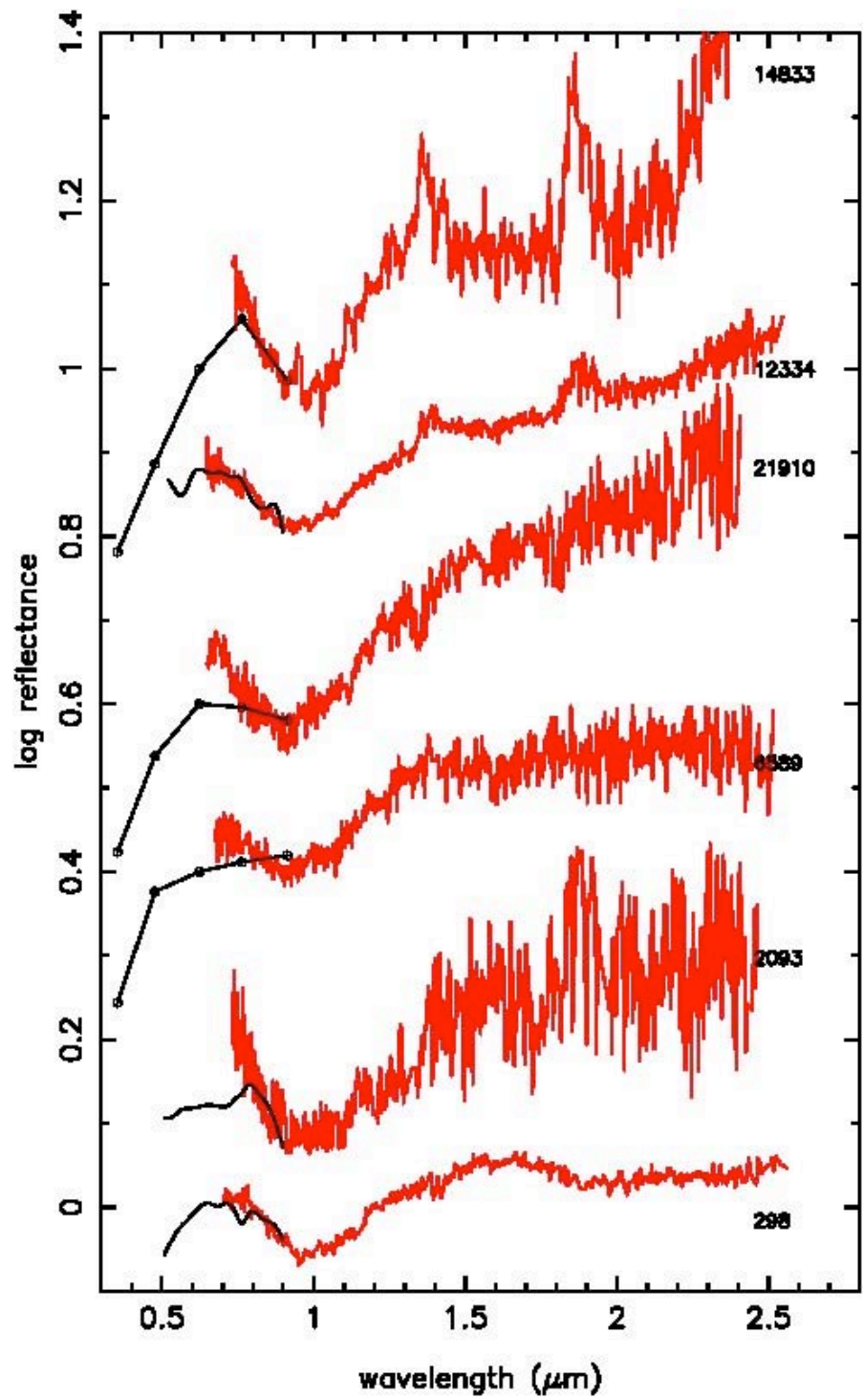

**Figure 12: Baptistina Asteroid Family**

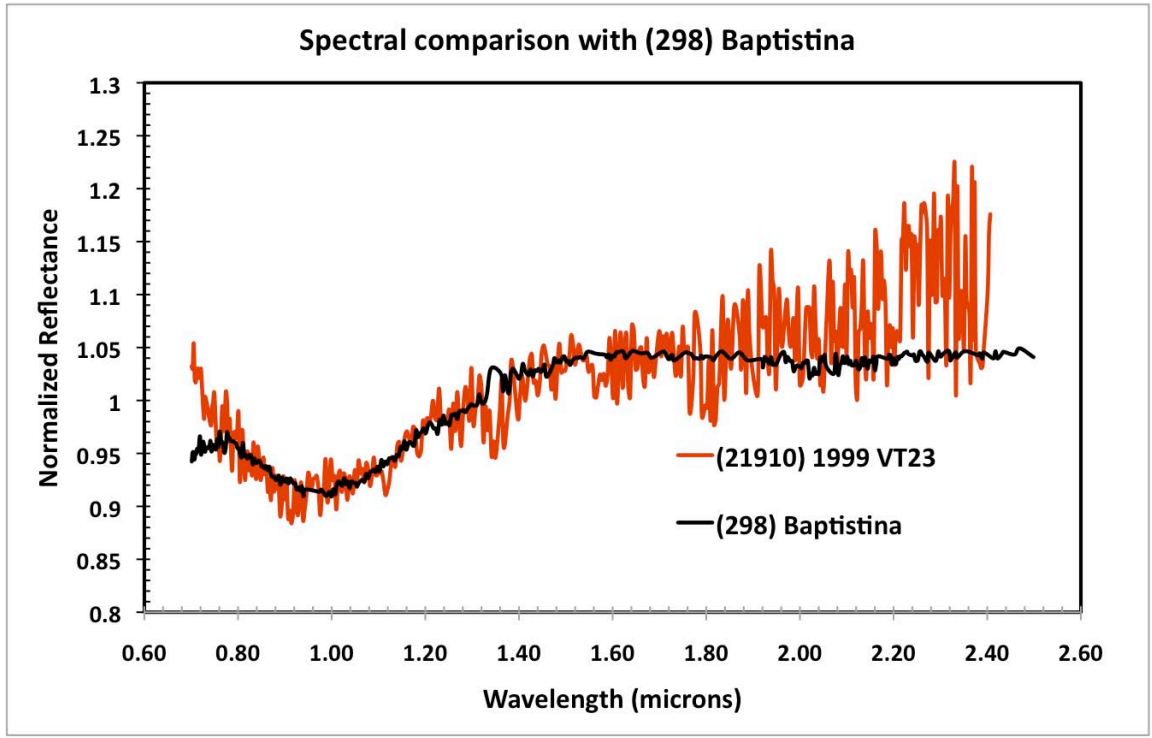

**Figure 13: Baptistina Asteroid Family**

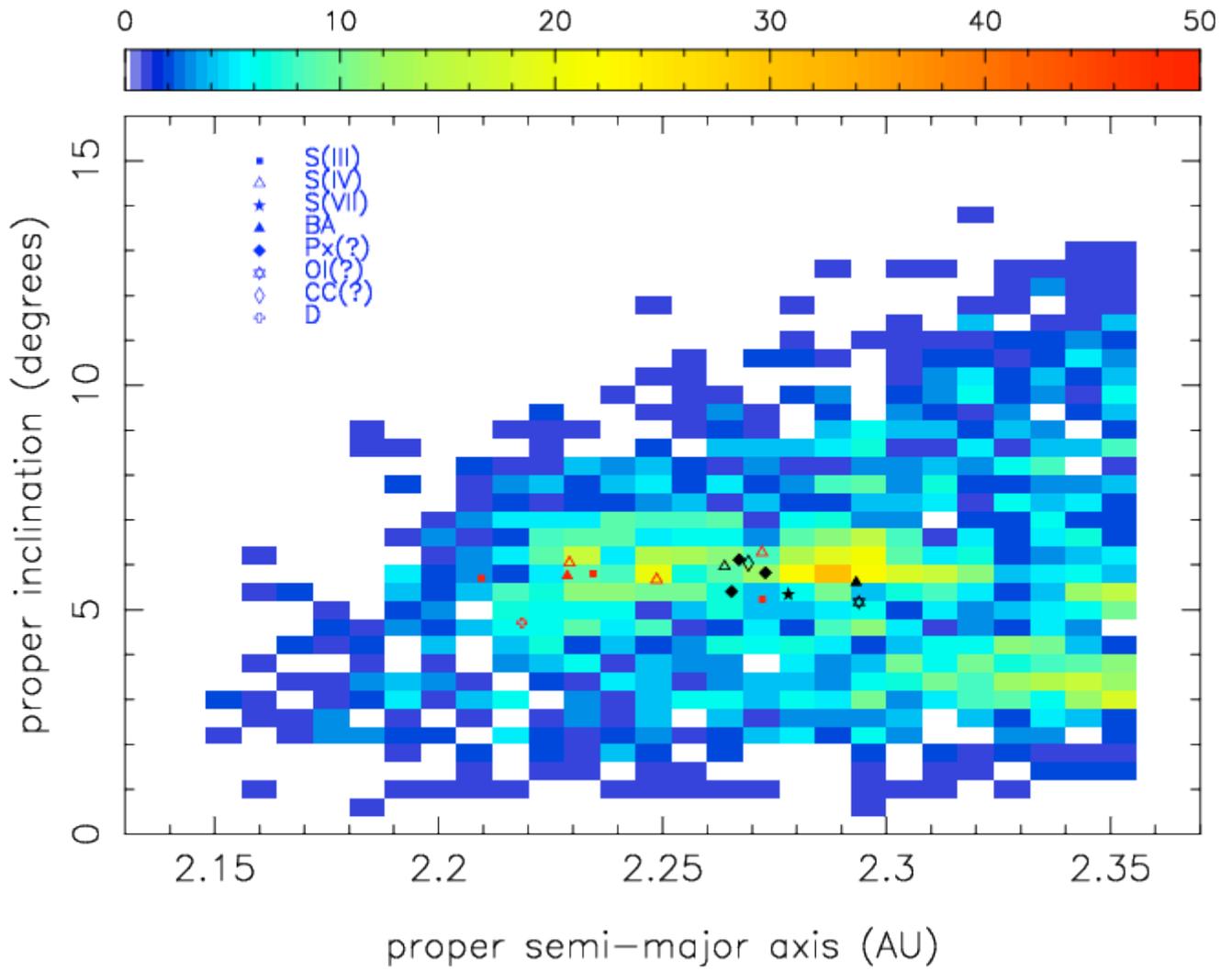

**Figure 14: Baptistina Asteroid Family**

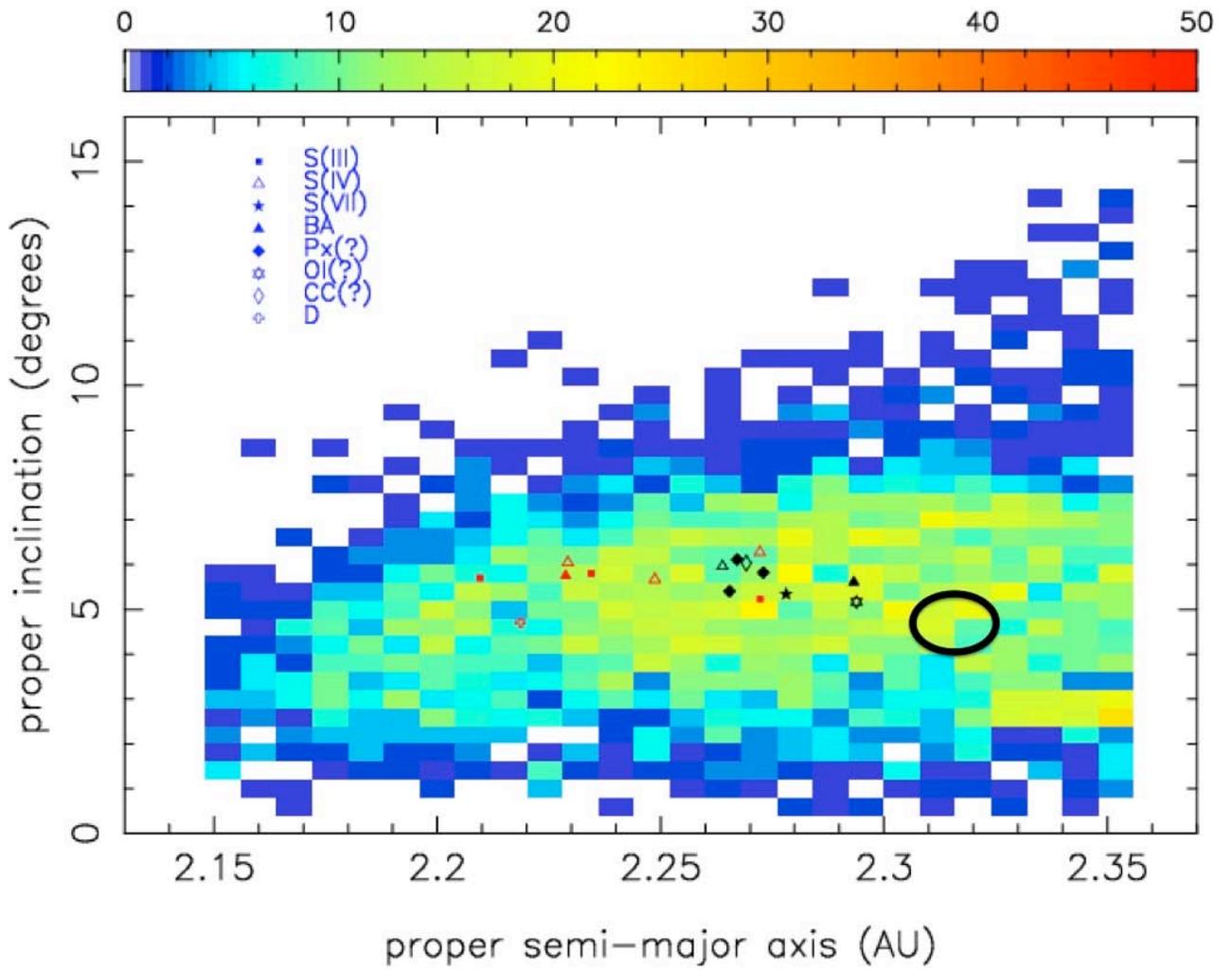

**Figure 15: Baptistina Asteroid Family**

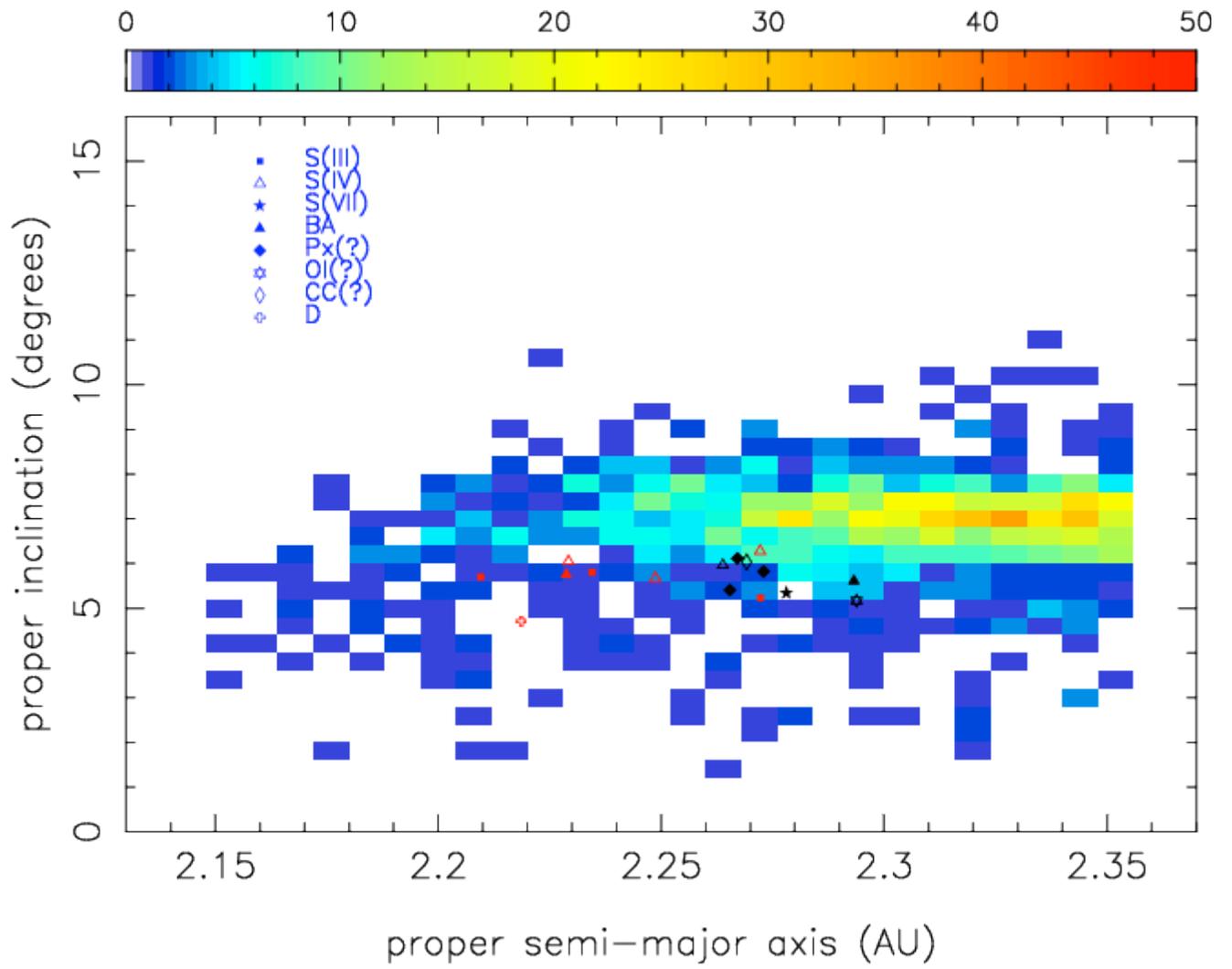